\documentclass[a4paper,fleqn,usenatbib,useAMS]{mnras}
\usepackage[T1]{fontenc}
\usepackage{ae,aecompl}
\usepackage{graphicx}
\usepackage{amsmath}
\usepackage{amssymb}
\usepackage{lineno}

\newcommand\xmm{{\it XMM-Newton}}

\newcommand\asca{{\it ASCA}}

\newcommand\ks{{\rm~ks}}

\newcommand\mpc{{\rm~Mpc}}

\newcommand\hz{{\rm~Hz}}
\newcommand\keV{{\rm~keV}}
\newcommand\ev{{\rm~eV}}

\title[Energy-dependent variability of Ark 120]{Energy-dependent variability of the bare Seyfert~1 galaxy Ark 120}
\author[L. Mallick et al.]{Labani Mallick$^{1}$\thanks{E-mail: labani@iucaa.in (LM)}, Gulab C. Dewangan$^{1}$, I. M. M$^{\rm c}$Hardy$^{2}$ and Mayukh Pahari$^{1}$ \\
$^1$ Inter-University Centre for Astronomy and Astrophysics, Post Bag 4, Ganeshkhind, Pune 411007, India\\
$^2$ School of Physics \& Astronomy, University of Southampton, Highfield, Southampton SO17 1BJ, UK}

\begin{document}

\pagerange{\pageref{firstpage}--\pageref{lastpage}} 

\maketitle

\label{firstpage}

\begin{abstract}
We present results from a detailed spectral-timing analysis of a long $\sim486$\ks{} \xmm{} observation of the bare Seyfert~1 galaxy Ark~120 which showed alternating diminution and increment in the 0.3$-$10\keV{} X-ray flux over four consecutive orbits in 2014. We study the energy-dependent variability of Ark~120 through broad-band X-ray spectroscopy, fractional root-mean-squared (rms) spectral modelling, hardness$-$intensity diagram and flux$-$flux analysis. The X-ray (0.3$-$10\keV{}) spectra are well fitted by a thermally Comptonized primary continuum with two (blurred and distant) reflection components and an optically thick, warm Comptonization component for the soft X-ray excess emission below $\sim2$\keV{}. During the first and third observations, the fractional X-ray variability amplitude decreases with energy while for second and fourth observations, X-ray variability spectra are found to be inverted-crescent and crescent shaped respectively. The rms variability spectra are well modelled by two constant reflection components, a soft excess component with variable luminosity and a variable intrinsic continuum with the normalization and spectral slope being correlated. The spectral softening of the source with both the soft excess and UV luminosities favour Comptonization models where the soft excess and primary X-ray emission are produced through Compton up-scattering of the UV and UV/soft X-ray seed photons in the putative warm and hot coronae, respectively. Our analyses imply that the observed energy-dependent variability of Ark~120 is most likely due to variations in the spectral shape and luminosity of the hot corona and to variations in the luminosity of the warm corona, both of which are driven by variations in the seed photon flux.
\end{abstract}

\begin{keywords}
black hole physics--galaxies: active--galaxies: X-rays--galaxies: individual: Ark~120
\end{keywords}

\section{Introduction}
Active galactic nuclei (AGN) emit radiation over the entire range of the electromagnetic spectrum where the X-rays are emitted from the innermost region of the accretion flow. The X-ray spectra of Seyfert~1 galaxies have the following main components: a power-law like primary emission with a high energy cut-off, `soft X-ray excess' emission below $\sim2$\keV{}, reflected emission consisting of a Fe~K line near 6\keV{} and a Compton hump in the $20-40$\keV{} energy range. It is widely accepted that Comptonization of lower energy seed photons is the driving mechanism for the production of the primary X-ray emission (e.g. \citealt{su80,ti94}). The seed photons are thought to arise  from the accretion disc with peak emission at optical/ultraviolet (UV) wavebands \citep{sh73}. However, the location and geometry of the scattering region are still a matter of debate. It has been suggested that the scattering region can be a hot corona above the cold disc \citep{ha91,ha93,po96}, a base of the relativistic jet \citep{fe99,fe04,ma05} or an advection-dominated accretion flow (ADAF; \citealt{na94, es97}). The radial extension of the corona is also unresolved: it could be either physically compact \citep{fa12,re13} or extended \citep{ku97,wi16,ch17}. In addition to the primary X-ray continuum, many Seyfert~1 galaxies show an excess emission below $\sim2$\keV{}, the origin of which is still controversial. Presently, there are two physical scenarios that can describe the origin of the soft X-ray excess emission -- thermal Comptonization of the optical/UV seed photons in an optically thick, warm corona \citep{ma98,de07,done12,lo12} and blurred reflection from an ionized accretion disc \citep{fa02, ro05, cr06, ga14}. The reflection features (Fe~K$_\alpha$ line and Compton hump) arise due to the photoelectric absorption followed by the fluorescence line emission below 10\keV{} and Compton scattering dominating above 10\keV{} in a dense and relatively cold medium like the accretion disc or torus. However, the complexity of the broad-band AGN spectra sometimes results in mean spectral model degeneracy \citep{de07, md17} and to overcome that problem, we need to study the variability properties of different spectral components.   

The X-ray emission from type~1 AGN is known to be strongly variable on a wide range of time-scales \citep{le99b,tu99,po12}. However, the exact nature of X-ray variability is not clearly understood. The X-ray variability may arise due to variations in the accretion rate (e.g. \citealt{are05}) and/or intrinsic luminosity of the X-ray emitting hot corona (e.g. \citealt{mar14,ma16}). Some of the observed variability can also be explained in terms of variations in the source height \citep{mf04}. One of the most promising techniques that can probe the variable behaviour of spectral components is the root-mean-squared (rms) variability spectrum since it connects the energy spectrum with the variability properties and has been proven to be very successful in explaining the variability of spectral components in AGN (e.g. MCG--6-30-15: \citealt{mi07}, 1H~0707--495: \citealt{fa12}, RX~J1633.3+4719: \citealt{ma16}, PG~1211+143: \citealt{lo16}). The rms spectrum determines the dynamical nature of distinct spectral components (Comptonization, reflection, absorption) or spectral parameters (normalization, spectral index etc.) and also the coupling between different components or parameters on various timescales.

In this work, we investigate the origin of energy-dependent variability of Ark~120 including the soft excess emission, the intrinsic coronal emission and UV/X-ray connection using the \xmm{} archival data from the 2014 observations. Ark~120 is a broad-line Seyfert~1 galaxy \citep{ve06} at a redshift of $z=0.0327$ \citep{th05} with Balmer lines full width at half-maximum FWHM(H$_\beta$) $\sim5800$\rm~km~s$^{-1}$ \citep{wa99}, an estimated black hole mass of $M_{\rm BH}\sim 1.5\times10^{8} M_{\odot}$ \citep{pe04} and low Eddington ratio of $L_{\rm bol}/L_{\rm E}\sim0.05$ \citep{vas07}. The source is well known for showing a strong soft excess \citep{ma14}, a deficit of intrinsic ultraviolet/X-ray ionized absorption \citep{cr01,va04} and no significant contamination from the host galaxy \citep{wa87} $-$ therefore considered as a `bare' nucleus and an ideal target to shed light on the central engine. Here we present the spectral-temporal analysis including the broad-band energy spectrum, flux$-$flux analysis, hardness$-$intensity diagram (HID), rms$-$flux relation and fractional rms spectrum to probe the variable nature of different spectral components in Ark~120.

\section{Observations and Data Reduction}
Ark~120 was observed with the \xmm{} telescope \citep{ja01} during four consecutive orbits (rev2614, rev2615, rev2616, rev2617) in 2014. We refer them 2014a, 2014b, 2014c and 2014d to represent their chronological order. Here we analyse the data from the European Photon Imaging Camera (EPIC-pn; \citealt{st01}) which was operated in the small window (\textsc{sw}) mode using the thin filter for all four observations. We processed the data sets with the Scientific Analysis System (\textsc{sas}~v.15.0.0) and used the most recent (as of 2016 September 4) calibration files. We have used unflagged events with \textsc{pattern}$\leq4$. We excluded the intervals of proton flares by creating a \textsc{gti} (Good Time Interval) file above 10\keV{} for the full field with the \textsc{rate}$<0.09$\rm~cts~s$^{-1}$, $0.11$\rm~cts~s$^{-1}$, $0.16$\rm~cts~s$^{-1}$ and $0.2$\rm~cts~s$^{-1}$ for 2014a, 2014b, 2014c and 2014d, respectively, to acquire the maximum signal-to-noise ratio. Then we extracted the source and background events with extraction radii of 30~arcsec and 60~arcsec, respectively. The \textsc{rmf} (Redistribution Matrix File) and \textsc{arf} (Ancillary Region File) for each EPIC-pn spectral data sets were produced with the \textsc{sas} tasks \textsc{rmfgen} and \textsc{arfgen}, respectively. We binned the spectra using the \textsc{grppha} tool with a minimum of 100 counts per bin. The source and background light curves for different energy bands were extracted using \textsc{xselect}~V.2.4~c from the cleaned pn events. Finally, we produced the background subtracted light curves with the \textsc{ftool} task \textsc{lcmath}. 

The Optical Monitor (OM; \citealt{ma01}) was operated in the imaging-fast (IF) mode using all six optical/UV filters (UVW2, UVM2, UVW1, U, B, V) with a total duration of $\sim96$\ks{} for each observation. There is a total of 80 OM exposures for each observation. We processed the fast mode OM data with the \textsc{sas} task \textsc{omfchain} and generated the optical/UV light curves. To process the imaging mode OM data, we used the \textsc{sas} task \textsc{omichain}. We detected the source in the sky aligned image for each filter. Then we checked the combined source list to find the right ascension, declination and background subtracted count rate of the source, corrected for coincidence losses. 

\begin{table*}
\caption{\xmm{}/EPIC-pn Observations of Ark~120 in 2014. Count rates for EPIC-pn are estimated in the 0.3$-$10\keV{} energy band.}
\scalebox{1.0}{%
\begin{tabular}{ccccccc}
\hline 
Revolution & Order No. & Obs. ID & Date & Filtered Duration & Net Exposure & Net Count Rate \\
           & &  & (yyyy-mm-dd)  & (\ks{}) & (\ks{}) & (counts~s$^{-1}$) \\                                      
\hline 
rev2614 & 2014a  & 0721600201 & 2014-03-18 & 121.0 & 65.71   & 28.3$\pm0.02$  \\ [0.2cm]
rev2615 & 2014b  & 0721600301 & 2014-03-20 & 122.0 & 73.02  & 23.7$\pm0.02$ \\ [0.2cm] 
rev2616 & 2014c  & 0721600401 & 2014-03-22 & 124.0 & 72.54  & 26.2$\pm0.02$  \\ [0.2cm]
rev2617 & 2014d  & 0721600501 & 2014-03-24 & 119.0 & 77.65  & 23.8$\pm0.02$ \\ [0.2cm]
\hline 
\end{tabular}}
\label{table1}           
\end{table*}

\begin{figure}
\includegraphics[scale=0.32,angle=-90]{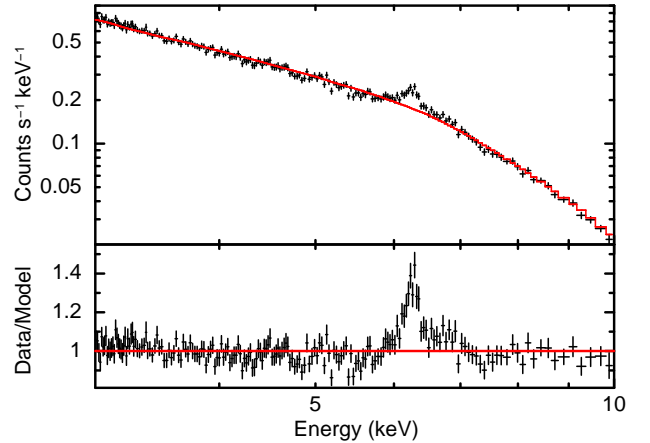}
\caption{The 2014a EPIC-pn spectral data, the absorbed power-law ($\Gamma = 1.77$) model (in solid blue) fitted in the 3$-$10\keV{} energy band and the data-to-model ratio which shows strong residuals in the Fe~K region. The spectrum is binned up for clarity.} 
\label{fig1a}
\end{figure}

\begin{figure}
\includegraphics[scale=0.32,angle=-90]{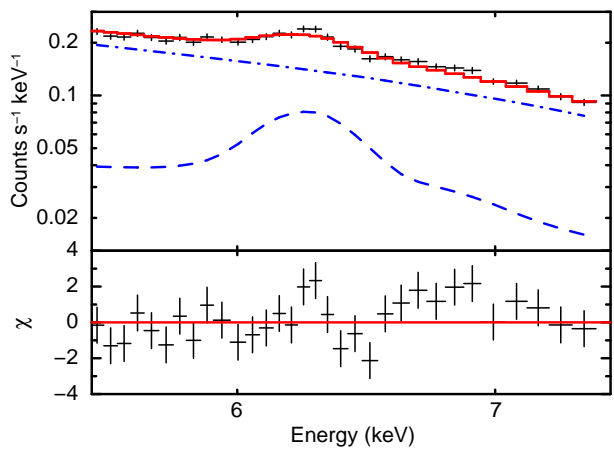}
\caption{The 2014a EPIC-pn $5.5-7.5$\keV{} spectral data, the blurred reflection model [\textsc{tbabs$\times$(relconv$\ast$xillver$+$cutoffpl)}] and the residuals. The spectrum is binned up for clarity.}
\label{fig1b}
\end{figure}

\begin{figure*}
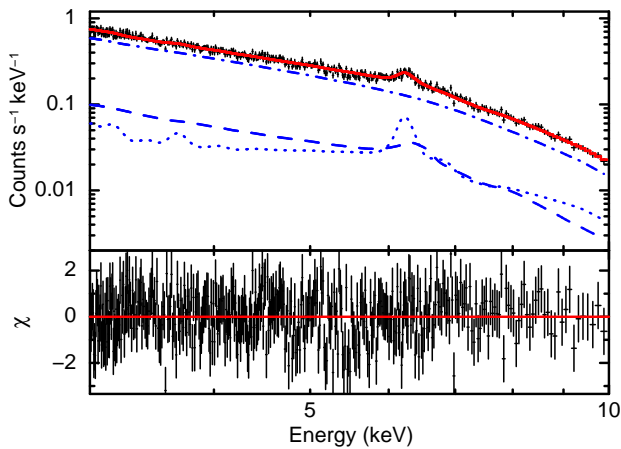

\includegraphics[scale=0.32,angle=-90]{fig1d.ps}
\includegraphics[scale=0.32,angle=-90]{fig1c.ps}
\caption{Left: The 2014a EPIC-pn hard band (3$-$10\keV{}) spectrum, the best-fitting model, 
[\textsc{tbabs$\times$(relconv$\ast$xillver+xillver+cutoffpl)}] and the residual spectrum. Right: The 2014a EPIC-pn $5.5-7.5$\keV{} spectral data, the blurred and unblurred reflection models and the residuals. The hard band best-fitting model has three main components: the primary emission with a high energy cutoff (in dash-dotted), highly ionized blurred disc reflection (in dashed line) and less ionized, distant reflection (in dotted line).}
\label{fig1dc}
\end{figure*}

\begin{figure}
\includegraphics[scale=0.32,angle=-90]{fig2.ps}
\caption{The 2014a broad-band (0.3$-$10\keV{}) EPIC-pn spectrum, the hard band best-fitting phenomenological model, \textsc{tbabs$\times$(zgauss1$+$zgauss2$+$zgauss3$+$zpowerlw)} (in solid red). The lower panel shows the ratio of the full band data to the hard band best-fitting model extrapolated down to 0.3\keV{}.}
\label{fig2}
\end{figure}

\begin{figure}
\includegraphics[scale=0.32,angle=-90]{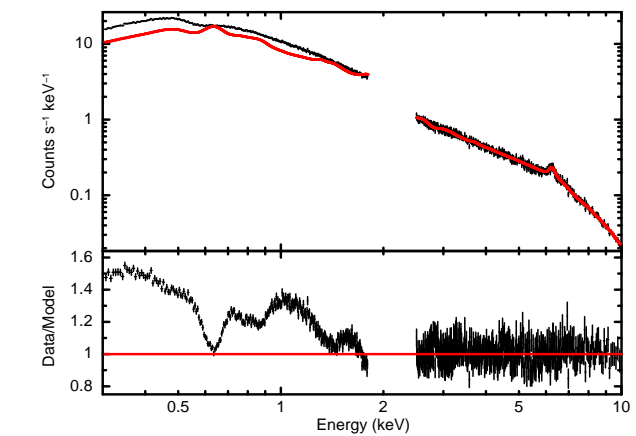}
\caption{The 2014a EPIC-pn 0.3$-$10\keV{} spectral data, the hard band best-fitting model, \textsc{tbabs$\times$(relconv$\ast$xillver+xillver+cutoffpl)} extrapolated to the lower energies and the data-to-model ratio.}
\label{fig3}
\end{figure}

\begin{figure*}
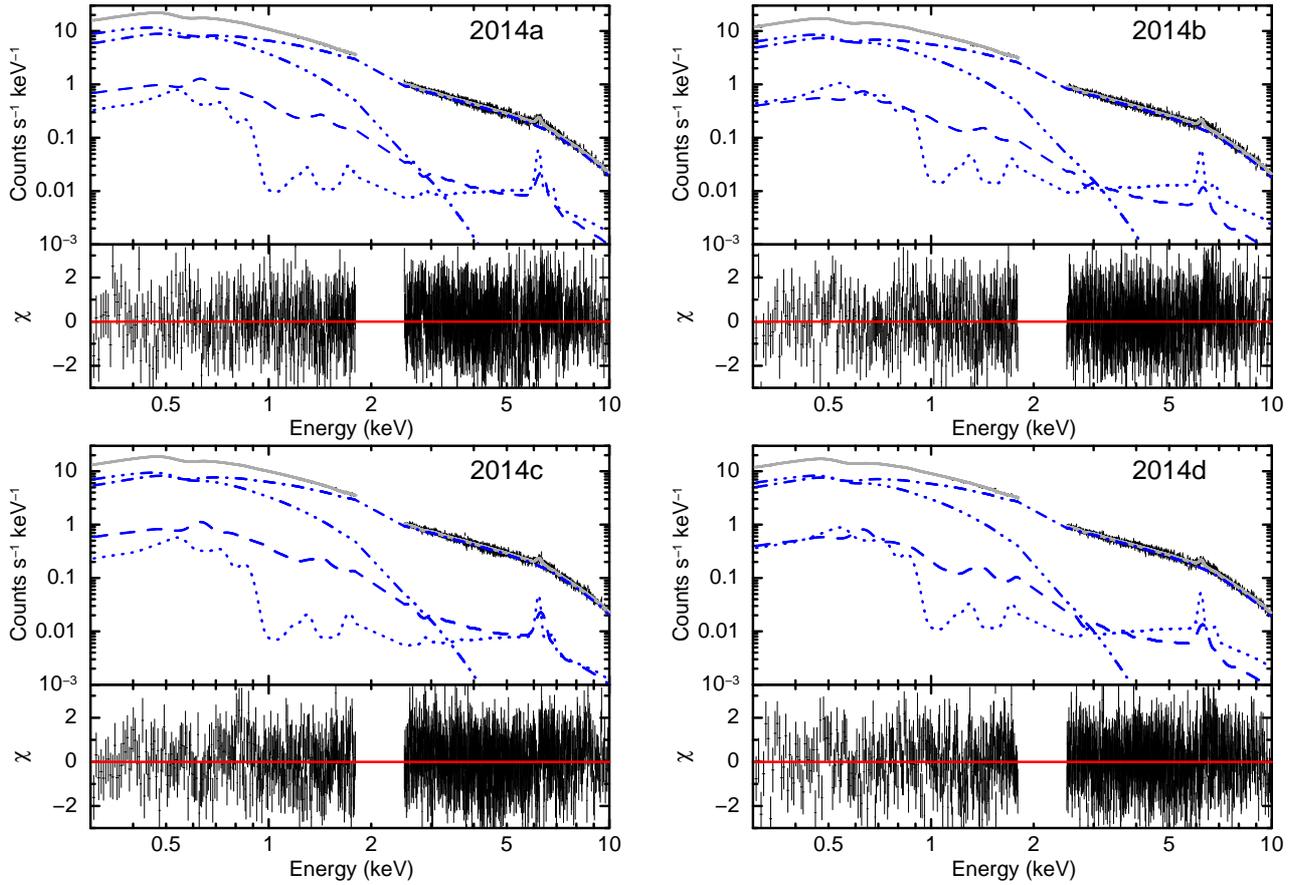

\includegraphics[scale=0.32,angle=-90]{fig4a.ps}
\includegraphics[scale=0.32,angle=-90]{fig4b.ps} 
\includegraphics[scale=0.32,angle=-90]{fig4c.ps}
\includegraphics[scale=0.32,angle=-90]{fig4d.ps}
\caption{The 2014 \xmm{}/EPIC-pn spectra and the deviations of the 0.3$-$10\keV{} observed data from the best-fitting model, \textsc{tbabs$\times$(optxagnf+relconv$\ast$xillver+xillver+nthcomp)}. The best-fitting model (solid line) consists of one thermally Comptonized primary continuum (dash-dotted line), intrinsic disc Comptonization for the soft excess (dash-dot-dot-dot line), more ionized blurred disc reflection (dashed line) and less ionized distant reflection (dotted line).}
\label{fig4}
\end{figure*}

\begin{figure*}
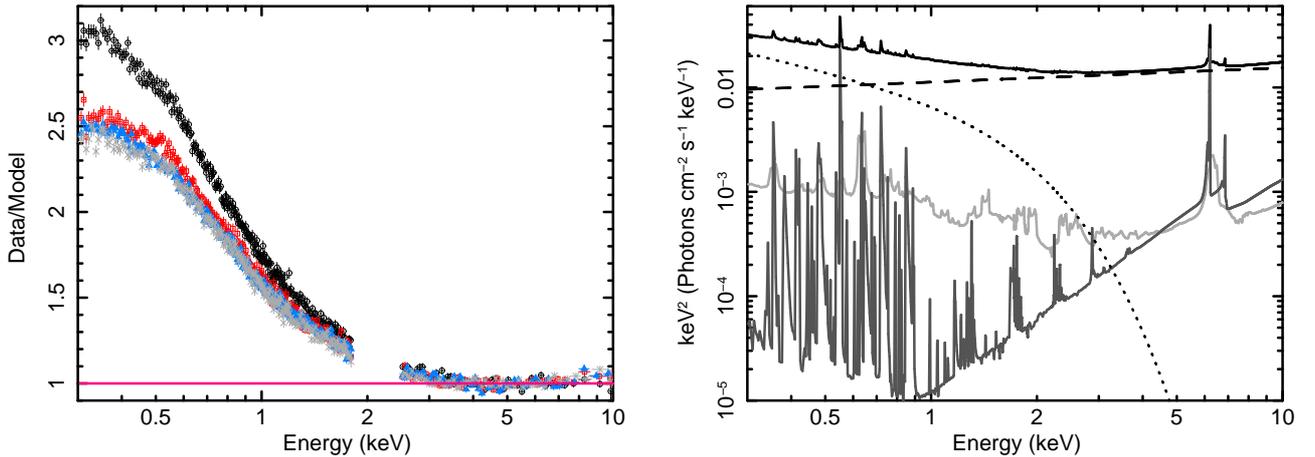

\includegraphics[scale=0.33,angle=-90]{fig5a.ps}
\includegraphics[scale=0.33,angle=-90]{fig5b.ps}
\caption{Left: Ratio of the 0.3$-$10\keV{} EPIC-pn spectral data sets and the hard band (3$-$10\keV{}) phenomenological model, \textsc{tbabs$\times$(zgauss1$+$zgauss2$+$zgauss3$+$zpowerlw)} ($\Gamma = 1.78$) extrapolated down to 0.3\keV{}. The circles, squares, triangles and crosses represent EPIC-pn data for the 2014a, 2014b, 2014c and 2014d observations, respectively. Right: The best-fitting physical model with all the model components excluding the Galactic absorption. The thermally Comptonized primary continuum (\textsc{nthcomp}) is shown in dashed black, the soft excess emission (\textsc{optxagnf}) is shown in dotted black, the disc reflection (\textsc{relconv$\ast$xillver}) is shown in light grey, the distant reflection (\textsc{xillver}) is shown in dark grey and the total model is shown in solid black.}
\label{fig5}
\end{figure*}

\section{Spectral Analysis}
We analysed the 2014 \xmm{}/EPIC-pn energy spectra of Ark~120 with \textsc{xspec}~v.12.8.2 \citep{ar96}. We employed the $\chi^{2}$ statistics to estimate the errors on the best-fitting model parameters at the 90~per~cent confidence limit which corresponds to $\Delta\chi^{2}$ = 2.706 for each parameter. We ignored the 1.8$-$2.5\keV{} band in order to avoid the effects of instrumental features on the broad-band energy spectrum (see e.g. \citealt{ma14,mar14}).

\subsection{The hard band EPIC-pn spectrum}
The X-ray energy spectrum of Ark~120 is complex in nature as evident from the previous 2013 observation \citep{ma14}. Therefore we began our analysis by fitting the 2014a hard band (3$-$10\keV{}) EPIC-pn spectrum with a power-law (\textsc{zpowerlw}) modified by the Galactic absorption. We modelled the Galactic absorption using \textsc{tbabs} in \textsc{xspec} with the interstellar Hydrogen column density fixed at $N_{\rm H}=9.78\times10^{20}$\rm~cm$^{-2}$ \citep{ka05}. The fit is unacceptable with $\chi^{2}$/d.o.f = 936/664. Figure~\ref{fig1a} shows the 2014a EPIC-pn 3$-$10\keV{} spectrum, absorbed power-law fit to the data and the data-to-model ratio plot with a strong residual in the Fe~K region. Initially, we fitted the Fe~K region with two narrow emission lines (\textsc{zgauss1+zgauss2}) by fixing the line widths at $\sigma=10$\ev{}, which improved the fit to $\chi^{2}$/d.o.f = 708/660 ($\Delta\chi^{2}$=$-$228 for 4 d.o.f). The lines are centred at $\sim6.45$\keV{} and $\sim6.96$\keV{}. However, a visual inspection of the residual indicates a broad emission feature at $\sim6.5$\keV{}. Therefore we added a third Gaussian line (\textsc{zgauss3}) which provided a statistically acceptable fit with $\chi^{2}$/d.o.f = 661.5/657 ($\Delta\chi^{2}$=$-$46.2 for 3 d.o.f). The best-fitting value of the broad emission line width is $\sigma_{B}=259^{+134}_{-80}$\ev{}. The line energies of two narrow and one broad emission lines are $E_{1}=6.44^{+0.02}_{-0.02}$\keV{}, $E_{2}=7.04^{+0.09}_{-0.06}$\keV{} and $E_{B}=6.5^{+0.14}_{-0.05}$\keV{} with equivalent widths of ${\rm EW_{1}}=41^{+17}_{-16}$\ev{}, ${\rm EW_{2}}=23^{+13}_{-14}$\ev{} and ${\rm EW_{B}}=103^{+27}_{-25}$\ev{}, respectively. We conclude that the EPIC-pn 3$-$10\keV{} spectrum is well fitted by a power-law with a photon index of $\Gamma=1.83\pm0.02$ along with one broad and two narrow iron emission lines. 

We then refitted the hard band (3$-$10\keV{}) spectrum with the physically motivated reflection models to understand the origin of the spectral lines. First, we used the relativistic reflection model \textsc{relconv$\ast$xillver} \citep{ga14} to obtain the reflected emission from the inner accretion disc. Since the \textsc{xillver} model assumes a power-law with a high energy cutoff as input continuum, we used a cutoff power-law (\textsc{cutoffpl}) model as the primary continuum. We consider a SMBH of spin $a=0$, disc inclination angle of $i=30^{\circ}$, an $\epsilon\propto r^{-3}$ emissivity law and an outer disc radius of $r_{\rm out}=400r_{\rm g}$ as assumed by \citet{na16}. The high energy cutoff of the primary Continuum was fixed at $E_{\rm cut}=1000$\keV{}. The fitting of the 3$-$10\keV{} spectrum with the blurred reflection model, \textsc{tbabs$\times$(relconv$\ast$xillver+cutoffpl)} provided a $\chi^{2}$/d.o.f = 677/660 with two residuals at around $\sim6.4$\keV{} and $\sim6.97$\keV{} (see Figure~\ref{fig1b}), which are in agreement with the results of \citet{na16}. The $6.4$\keV{} emission represents the Fe~K$_{\alpha}$ line, while the $6.97$\keV{} emission feature most likely arises from a blend of the Fe~XXVI Ly${\alpha}$ line at $\sim6.9$\keV{} and Fe~K$_{\beta}$ line at $\sim7.06$\keV{}. In order to model these two narrow features, we added one distant reflection (\textsc{xillver}) model which improved the fit to $\chi^{2}$/d.o.f = 661.2/658 ($\Delta\chi^{2}$=$-$15.7 for 2 d.o.f) with no significant residuals. The 2014a EPIC-pn 3$-$10\keV{} spectral data, the best-fitting model, \textsc{tbabs$\times$(relconv$\ast$xillver+xillver+cutoffpl)} and the residuals are shown in Figure~\ref{fig1dc} (left). The accretion disc inner radius and ionization parameter obtained from the hard band spectral fitting are $r_{\rm in}=66.8^{+33.2p}_{-56.3}r_{\rm g}$ and $\xi=986^{+132}_{-349}$~erg~cm~s$^{-1}$, respectively. Thus the modelling of the hard band spectrum suggests the presence of a highly ionized blurred disc reflection and a weakly ionized distant reflection in Ark~120. Fig.~\ref{fig1dc} (right) shows the Fe~K (5.5$-$7.5\keV{}) region spectral data with all the model components. 

\begin{table*}
 \centering
 \caption{The best-fitting spectral model parameters for the 2014 \xmm{} observations (0.3$-$10\keV{}) of Ark~120. Parameters with notations `(f)' and $\ast$ indicate fixed and tied values, respectively.}
\begin{center}
\scalebox{0.95}{%
\begin{tabular}{cccccc}
\hline 
Model & \textsc{tbabs$\times$(optxagnf+relconv$\ast$xillver+xillver+nthcomp)} & 2014a & 2014b & 2014c & 2014d  \\
\hline
 &Date& 2014-03-18 & 2014-03-20  & 2014-03-22  & 2014-03-24  \\                                       
\hline
 & Obs. ID & 0721600201 & 0721600301 & 0721600401 & 0721600501 \\ 
\hline
Component & Parameter &    \\
\hline 
\textsc{tbabs} & $N_{\rm H}$(10$^{20}$ cm$^{-2}$) & 9.78(f) & 9.78(f) & 9.78(f) & 9.78(f) \\
\hline
\textsc{optxagnf} & $\log(L_{\rm s}/L_{\rm E}$) & $-0.9^{+0.3}_{-0.2}$ & $-1.26^{+0.24}_{-0.15}$ & $-0.95^{+0.26}_{-0.39}$ & $-1.28^{+0.33}_{-0.17}$ \\ [0.2cm]
             & $a$ & 0$^{\ast}$ & 0$^{\ast}$ & 0$^{\ast}$ & 0$^{\ast}$  \\ [0.2cm]             
       & $T_{\rm SE}$(\keV{}) & $0.38^{+0.04}_{-0.09}$ & $0.38^{+0.02}_{-0.02}$ & $0.38^{+0.03}_{-0.03}$ & 0.34$^{+0.02}_{-0.02}$ \\ [0.2cm]       
       & $\tau$ & 11.4$^{+2.2}_{-0.7}$ & 12.1$^{+0.6}_{-0.5}$ & 11.8$^{+0.5}_{-0.5}$ & 13.2$^{+0.6}_{-0.5}$ \\ [0.2 cm]       
 & $r_{\rm corona}$($r_{\rm g}$) & $>8.1$ & $>8.1$ & $>7.6$ & $>10.9$ \\ [0.2 cm] 
 & $r_{\rm out}$($r_{\rm g}$) & 400(f) & 400(f) & 400(f) & 400(f) \\ [0.2cm] 
 & $f_{\rm pl}$ & 0(f) & 0(f) & 0(f) & 0(f)  \\ [0.2cm] 
\hline
\textsc{relconv} & $q$ & 3(f) & 3(f) & 3(f) & 3(f)\\ [0.2cm] 
                 & $a$ & 0(f) & 0(f) & 0(f) & 0(f)  \\ [0.2cm]  
                 & $i$(deg) & 30(f) & 30(f) & 30(f) & 30(f) \\ [0.2cm]
 & $r_{\rm in}$($r_{\rm g}$) & $>38.8$ & $>25.1$ & $>30.1$ & $>12.7$ \\ [0.2 cm]
  & $r_{\rm out}$($r_{\rm g}$) & 400(f) & 400(f) & 400(f) & 400(f) \\ [0.2cm] 

\textsc{xillver} & $A_{\rm Fe}$ & 1.2$^{+0.6}_{-0.3}$ & 0.8$^{+0.1}_{-0.1}$ & 1.3$^{+0.5}_{-0.4}$ & 0.8$^{+0.1}_{-0.1}$\\ [0.2cm]      
       & $\xi_{1}$(erg~cm~s$^{-1}$) & 497$^{+37}_{-188}$ & 473$^{+41}_{-237}$ & 454$^{+52}_{-183}$ & 269$^{+236}_{-58}$ \\ [0.2cm]       
       & $\Gamma$ & 1.87$^{\ast}$ & 1.83$^{\ast}$ & 1.82$^{\ast}$ & 1.82$^{\ast}$ \\ [0.2cm]       
       & $N_{\rm xillver1}$(10$^{-8}$) & 3.7$^{+2.4}_{-0.9}$ & 2.4$^{+3.5}_{-0.9}$ & 4.4$^{+3.6}_{-1.3}$ & 5.1$^{+3.0}_{-3.4}$ \\ [0.2cm]       
       & $i$(deg) & 30$^{\ast}$ & 30$^{\ast}$ & 30$^{\ast}$ & 30$^{\ast}$ \\ [0.2cm]     
       & $E_{\rm cut}$(\keV{}) & 1000(f) & 1000(f) & 1000(f) & 1000(f)  \\ [0.2cm] 
\hline       
\textsc{xillver} & $A_{\rm Fe}$ & 1.2$^{\ast}$ & 0.8$^{\ast}$ & 1.3$^{\ast}$ & 0.8$^{\ast}$ \\ [0.2cm]      
       & $\xi_{2}$(erg~cm~s$^{-1}$) & 12$^{+9}_{-10}$ & 20$^{+1}_{-9}$ & 20$^{+3}_{-7}$ & 20$^{+3}_{-9}$ \\ [0.2cm]       
       & $\Gamma$ & 1.87$^{\ast}$ & 1.83$^{\ast}$ & 1.82$^{\ast}$ & 1.82$^{\ast}$ \\ [0.2cm]       
       & $N_{\rm xillver2}$(10$^{-6}$) & 3.8$^{+5.0}_{-1.9}$ & 3.0$^{+2.4}_{-0.6}$ & 1.9$^{+1.2}_{-0.3}$ & 2.7$^{+2.8}_{-0.6}$ \\ [0.2cm]       
       & $i$(deg) & 30$^{\ast}$ & 30$^{\ast}$ & 30$^{\ast}$ & 30$^{\ast}$ \\ [0.2cm]     
       & $E_{\rm cut}$(\keV{}) & 1000$^{\ast}$ & 1000$^{\ast}$ & 1000$^{\ast}$ & 1000$^{\ast}$  \\ [0.2cm] 
\hline
\textsc{nthcomp} & $\Gamma$ & 1.87$^{+0.02}_{-0.03}$ & 1.83$^{+0.02}_{-0.02}$ & 1.82$^{+0.01}_{-0.02}$ & 1.82$^{+0.02}_{-0.01}$  \\ [0.2cm]
       & $T_{\rm e}$(\keV{}) & 350(f) & 350(f) & 350(f) & 350(f) \\ [0.2cm]
       & $T_{\rm bb}$(\ev{}) & 20(f) & 20(f) & 20(f) & 20(f) \\ [0.2cm]
       & $N_{\rm nth}$(10$^{-3}$)$^{a}$ & 11.2$^{+0.3}_{-0.4}$ & 9.6$^{+0.3}_{-0.3}$ & 10.7$^{+0.3}_{-0.3}$ & 9.9$^{+0.3}_{-0.2}$ \\ [0.2cm]  
\hline
\textsc{flux} & $F_{0.3-2}$(10$^{-11}$)$^{b}$ & 3.6 & 3.0 & 3.3 & 3.0 \\ [0.2cm]   
     & $F_{2-10}$(10$^{-11}$)$^{b}$ & 3.9 & 3.5 & 4.0 & 3.7 \\ [0.2cm]        
     & $\chi^2$/$\nu$ & 1105/1056 & 1141.2/1062 & 1119.4/1103 & 1158.6/1101 \\ [0.2cm]                   

\hline
\end{tabular}}
\end{center}
Notes:$^a$~Normalization in units of photons~cm$^{-2}$~s$^{-1}$~keV$^{-1}$ at 1\keV{}. $^b$~Observed flux in units of erg~cm$^{-2}$~s$^{-1}$.
\label{table2}
\end{table*}

\subsection{The broad-band EPIC-pn spectrum}
To study the soft X-ray excess emission from Ark~120, we extrapolated our hard band best-fitting phenomenological model, \textsc{tbabs$\times$(zgauss1$+$zgauss2$+$zgauss3$+$zpowerlw)} down to 0.3\keV{}, which revealed a strong soft excess emission below 2\keV{} (see Figure~\ref{fig2}). The soft excess is very similar to that found in other Seyfert~1 galaxies and QSOs \citep{cz03,gi04,cr06,pa07}. Although one blurred and one distant reflected emission successfully explained the 3$-$10\keV{} X-ray spectrum, the extrapolation of the hard band best-fitting physical model \textsc{tbabs$\times$(relconv$\ast$xillver+xillver+cutoffpl)} significantly underestimates the flux at lower energies with $\chi^{2}$/d.o.f = 51399/1068 (see Figure~\ref{fig3}). The fitting of the broad-band (0.3$-$10\keV{}) pn data with the reflection models resulted in a higher value of the spectral index, $\Gamma\sim2.3$ and a poor fit with $\chi^{2}$/d.o.f = 2080/1060. Therefore we added intrinsic disc Comptonization model \textsc{optxagnf} (\citealt{done12}) following the approach of \citet{ma14} in order to account for the soft excess emission only. The \textsc{optxagnf} is an energetically self-consistent model as it assumes that the gravitational energy released at each radius in the accretion disc is emitted as a blackbody radiation at radii larger than a certain radius, known as the coronal radius ($r_{\rm corona}$). However, at radii less than $r_{\rm corona}$, the energy is partitioned between powering the soft X-ray excess and hard X-ray power-law with a cutoff energy of $100$\keV{}. The lower limit of the previously non-detected high energy cutoff of Ark~120 is $190$\keV{} \citep{ma14}, which is higher than the cutoff energy assumed by \textsc{optxagnf} model. Hence we switched off the hard power-law component of the \textsc{optxagnf} model and modelled only the soft excess emission by setting $f_{\rm pl}=0$, which measures the fraction of the power emitted in the hard Comptonization component below $r_{\rm corona}$. The normalization of the model is determined by four parameters: black hole mass ($M_{\rm BH}$), dimensionless spin ($a$), proper distance ($d$) and soft X-ray luminosity in units of Eddington luminosity ($L_{\rm s}/L_{E}$) and therefore the model requires the normalization to be fixed at 1. We fixed the black hole mass, proper distance and outer disc radius at $1.5\times10^8 M_{\rm \odot}$ \citep{pe04}, 134\mpc{} and 400$r_{\rm g}$ respectively. In \textsc{xspec}, the 0.3$-$10\keV{} model reads as {\small\textsc{tbabs$\times$(optxagnf+relconv$\ast$xillver+xillver+cutoffpl)}}, which resulted in a statistically acceptable fit with $\chi^{2}$/d.o.f = 1105.1/1056 = 1.04. For physical consistency, we replaced the \textsc{cutoffpl} model for the primary emission by a thermally Comptonized continuum model \textsc{nthcomp} \citep{zd96}. We assumed an electron temperature of $kT_{\rm e}=350$\keV{} which agrees with the high energy cutoff of $E_{\rm cut}\simeq1000$\keV{}, since $E_{\rm cut}$ is about a factor of 2 or 3 higher than the hot coronal temperature. The disc blackbody seed photon temperature was fixed at $kT_{\rm bb}=20$\ev{}, which is the maximum disc temperature relevant for Ark~120, given its mass and accretion rate \citep{fr02}. The fitting of the broad-band (0.3$-$10\keV{}) data with the {\small\textsc{tbabs$\times$(optxagnf+relconv$\ast$xillver+xillver+nthcomp)}} model provided the similar quality fit with $\chi^{2}$/d.o.f = 1105/1056 = 1.04. The 2014 individual EPIC-pn spectrum, the best-fitting physical model with components and the deviations of the broad-band pn data from the best-fitting model are shown in Figure~\ref{fig4}. The best-fitting model parameters for all four 2014 observations (2014a, 2014b, 2014c and 2014d) are listed in Table~\ref{table2}. Our broad-band spectral modelling suggests that the soft excess emission from Ark~120 is produced through thermal Comptonization of the disc seed photons in a warm corona with an electron temperature of $kT_{\rm SE}\sim0.36$\keV{} and an optical depth of $\tau\sim12$.


\begin{figure*}
\includegraphics[width=1.8\columnwidth,angle=-0]{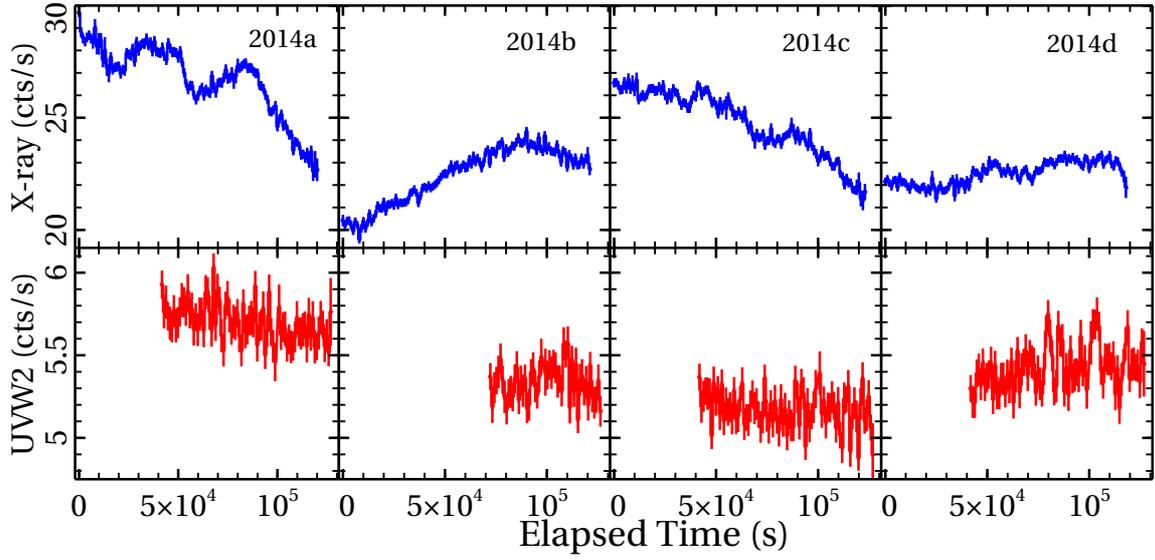}
\caption{The background subtracted EPIC-pn ($0.3-10$\keV{}) and UVW2 light curves (in 1\ks{} time bins) of Ark~120.}  
\label{lc}
\end{figure*}

\subsection{Joint Analysis of Four EPIC-pn Spectra}
We then performed the joint fitting of all four EPIC-pn spectra and studied the variability of spectral components. First, we applied the 3$-$10\keV{} best-fitting phenomenological model, \textsc{tbabs$\times$(zgauss1$+$zgauss2$+$zgauss3$+$zpowerlw)} to all four hard band EPIC-pn data. We tied all the parameters except the power-law normalization which resulted in a statistically acceptable fit with $\chi^{2}$/d.o.f = 2730.7/2750. As before, we extrapolated the hard band best-fitting phenomenological model to lower energies. The ratio of the broad-band pn data to the extrapolated hard band best-fitting phenomenological model is shown in Figure~\ref{fig5}~(left). Interestingly, we found that the strength of the soft excess component is variable between observations. The soft component has the highest flux level for 2014a, moderate for 2014b and lowest during the 2014c and 2014d observations. In order to probe the soft excess variability over the observed $\sim1$-week period, we applied the broad-band (0.3$-$10\keV{}) best-fitting physical model, {\small\textsc{tbabs$\times$(optxagnf+relconv$\ast$xillver+xillver+nthcomp)}} to all four EPIC-pn spectral data sets. Initially, all the parameters were tied except the normalization of the primary continuum (\textsc{nthcomp}) which resulted in a statistically poor fit with $\chi^{2}$/d.o.f = 9475.4/4355. Then we allowed the soft excess luminosity, $L_{\rm s}/L_{E}$ of the \textsc{optxagnf} model to vary between observations which improved the fit to $\chi^{2}$/d.o.f = 4750.7/4352 = 1.1. If we left the spectral index of the primary continuum to vary, we find a best-fit of $\chi^{2}$/d.o.f = 4665.6/4349 = 1.07 ($\Delta\chi^{2}$=$-$85 for 3 d.o.f) with no significant residual. The best-fitting values of the coronal radius, ionization parameters of the blurred and distant reflection components are $r_{\rm corona}=7.6^{+0.1}_{-0.1}r_{\rm g}$, $\xi_{1}=500^{+3}_{-149}$~erg~cm~s$^{-1}$ and $\xi_{2}=20^{+1}_{-2}$~erg~cm~s$^{-1}$, respectively. The best-fitting physical model along with the model components are shown in Fig.~\ref{fig5}~(right). The joint fitting of all four EPIC-pn spectral data sets indicate that the observed X-ray spectral variability of Ark~120 is due to variations in the normalization and photon index of the primary X-ray emission as well as the luminosity of the soft excess emission.

\begin{figure*}
\includegraphics[width=1.8\columnwidth]{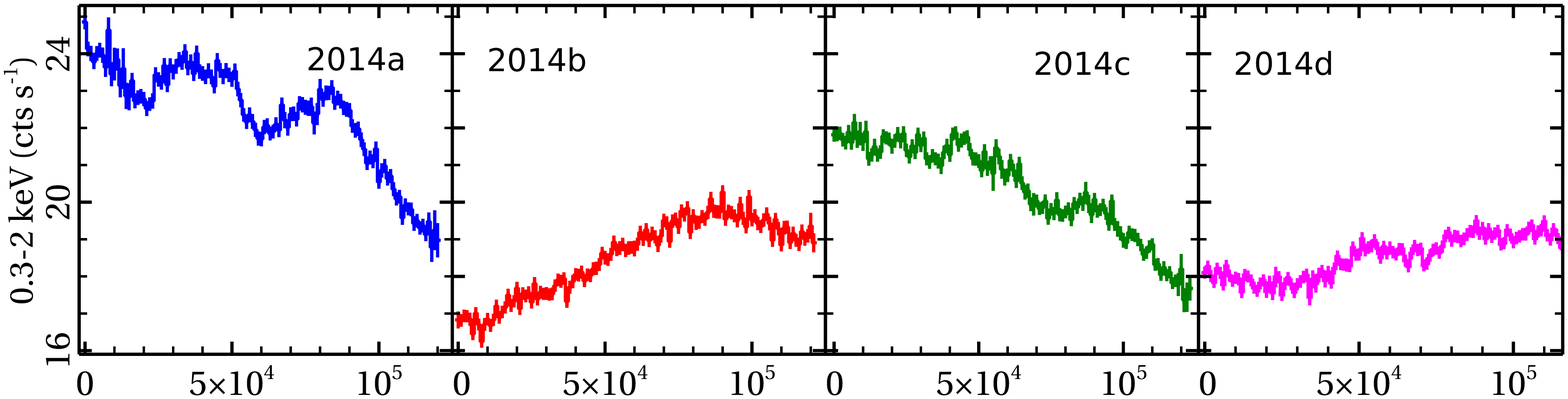}
\includegraphics[width=1.8\columnwidth]{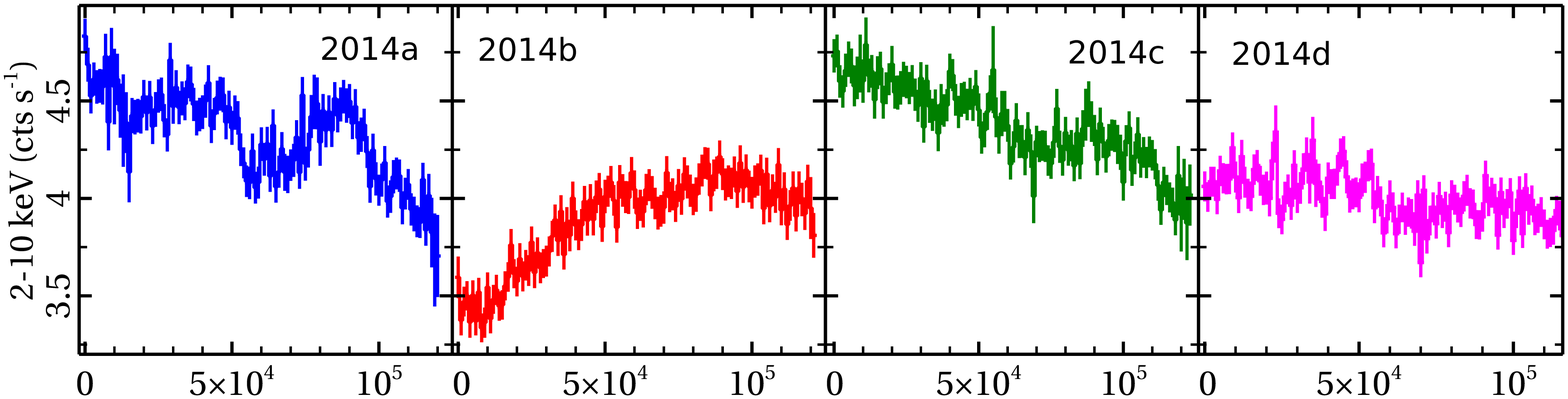}
\includegraphics[width=1.8\columnwidth]{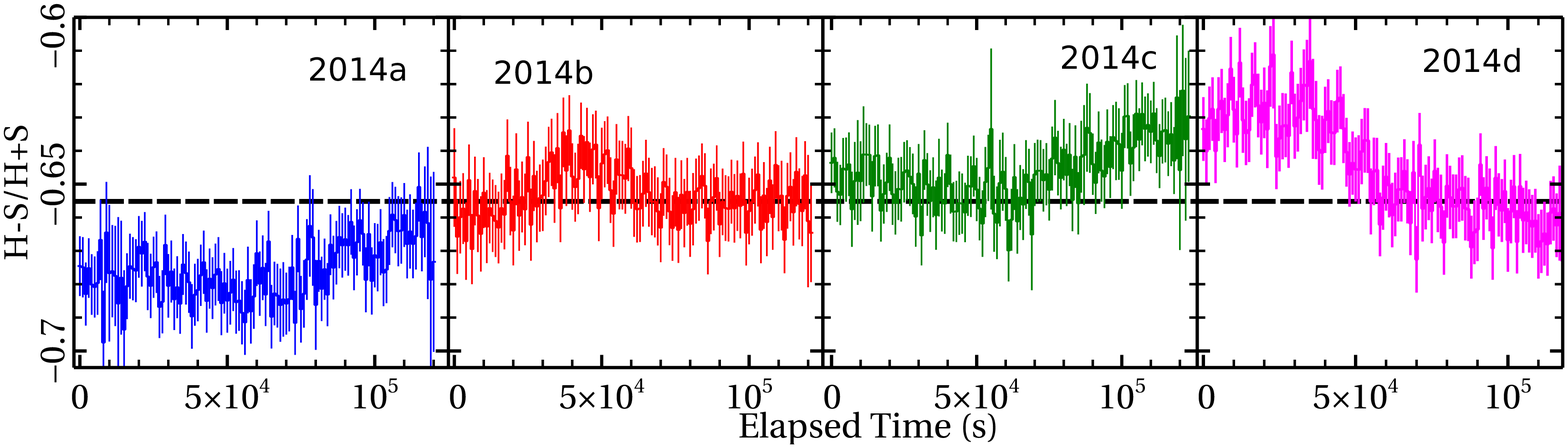}
\caption{The EPIC-pn light curves of Ark~120 in the soft (S=0.3$-$2\keV{}, top panel) and hard (H=2$-$10\keV{}, middle panel) X-ray bands. The bottom panel shows the hardness ratio, H$-$S/H+S. The dashed black line corresponds to the constant hardness ratio of $-0.6552$. The time resolution in each panel is 1\ks{}/bin.}
\label{lc_hr}
\end{figure*}

\begin{table*}
\centering
\caption{The mean count rate, absolute rms amplitude and fractional rms variability amplitude ($F_{\rm var}$) in different energy bands for four EPIC-pn observations.}
\begin{center}
\scalebox{1.0}{%
\begin{tabular}{ccccccc}
\hline 
Obs. Id & Order No. & Energy range & Mean (counts~s$^{-1}$) & rms (counts~s$^{-1}$) & $F_{\rm var}$(per~cent) \\                                   
\hline 
0721600201 & 2014a & $0.3-10\keV{}$ & 26.7$\pm0.25$  & 1.58$\pm0.02$  &5.91$\pm0.08$  \\ [0.2cm]
           &       & $0.3-2\keV{}$  & 22.37$\pm0.23$ & 1.38$\pm0.02$  & 6.16$\pm0.09$  \\ [0.2cm]
           &       & $2-10\keV{}$   & 4.32$\pm0.1$  &  0.21$\pm0.01$  & 4.77$\pm0.22$ \\ [0.2cm]
\hline             
0721600301 &2014b  & $0.3-10\keV{}$ & 22.42$\pm0.2$  & 1.22$\pm0.02$  &5.46$\pm0.08$  \\ [0.2cm]
           &       & $0.3-2\keV{}$  & 18.53$\pm0.18$ & 1.02$\pm0.02$  & 5.48$\pm0.09$ \\ [0.2cm]
           &       & $2-10\keV{}$   & 3.89$\pm0.08$  & 0.21$\pm0.01$  & 5.49$\pm0.2$ \\[0.2cm]
\hline                 
0721600401 & 2014c & $0.3-10\keV{}$ &24.76$\pm0.22$ & 1.42$\pm0.02$   &5.74$\pm0.08$   \\ [0.2cm]
           &       & $0.3-2\keV{}$  &20.38$\pm0.2$  & 1.24$\pm0.02$   &6.07$\pm0.09$   \\ [0.2cm]
           &       & $2-10\keV{}$   &4.38$\pm0.09$  & 0.18$\pm0.01$   & 4.21$\pm0.21$  \\[0.2cm]
\hline             
0721600501 & 2014d & $0.3-10\keV{}$ &22.52$\pm0.19$ & 0.46$\pm0.02$   &2.04$\pm0.08$ \\ [0.2cm]
           &       & $0.3-2\keV{}$  &18.52$\pm0.18$ & 0.52$\pm0.02$   & 2.8$\pm0.09$  \\ [0.2cm]
           &       & $2-10\keV{}$   &4.0$\pm0.08$   & 0.1$\pm0.01$   &2.4$\pm0.22$ \\ [0.2cm]          
\hline 
\end{tabular}}
\label{table3}
\end{center}          
\end{table*}

\begin{figure*}
\includegraphics[width=2\columnwidth]{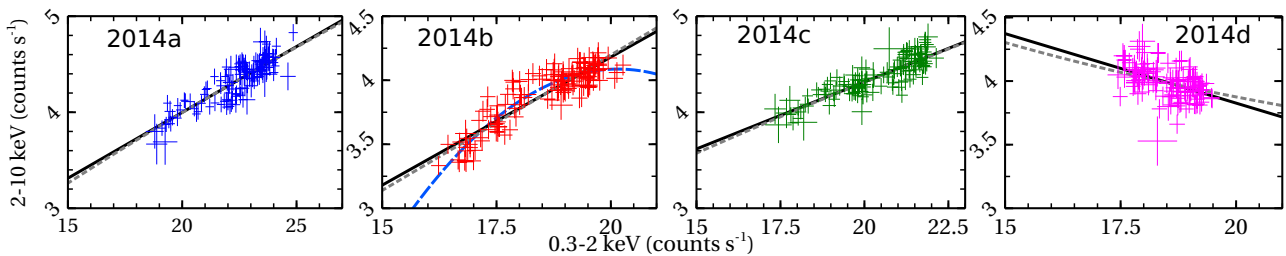}
\caption{The EPIC-pn count rate in the 2$-$10\keV{} band is plotted against the count rate in the 0.3$-$2\keV{} band at a resolution of 1\ks{}/bin. The linear and power-law fits to the data are represented by solid black and dotted grey lines, respectively. Both linear and power-law models provide the similar good quality fit except for 2014b where a quadratic model (in dashed blue line) provides a better fit to the data. Also for the 2014d observation, an anti-correlation between soft and hard band count rates is clearly seen.}
\label{flux_flux}
\end{figure*}

\begin{table*}
\centering
\caption{The parameters obtained from the linear and power-law fitting to the Flux$-$Flux plots.}
\begin{center}
\scalebox{0.9}{%
\begin{tabular}{ccccccccc}                        
\hline
Order No.   & & linear model ($\rm H=m\times\rm S+c$) & & & power-law model ($\rm H=\alpha \times\rm S^{\beta}$) & \\ 
\hline       
                 & $m$ & $c$ & $\chi^2$/d.o.f & $\alpha$ & $\beta$  & $\chi^2$/d.o.f \\ [0.2cm]
\hline 
 2014a & 0.14$\pm0.01$    & 1.23$\pm0.22$ & 187.2/119 & 0.47$\pm0.07$  & 0.71$\pm0.05$ & 188.1/119  \\ [0.2cm]
 2014b & 0.2$\pm0.01$     & 0.15$\pm0.21$ & 197.1/120 & 0.24$\pm0.04$  & 0.95$\pm0.05$ & 196.7/120  \\ [0.2cm]
 2014c & 0.14$\pm0.01$    & 1.5$\pm0.22$  & 155.3/122 & 0.61$\pm0.09$  & 0.65$\pm0.05$ & 155.8/122  \\ [0.2cm]
 2014d & $-$0.11$\pm0.02$ & 5.97$\pm0.41$ & 193.3/117 & 16.8$\pm4.2$   & $-$0.49$\pm0.1$ & 193.1/117  \\ [0.2cm]
\hline 
\end{tabular}}
\end{center}
\label{table4}           
\end{table*}

\begin{figure}
\includegraphics[width=\columnwidth]{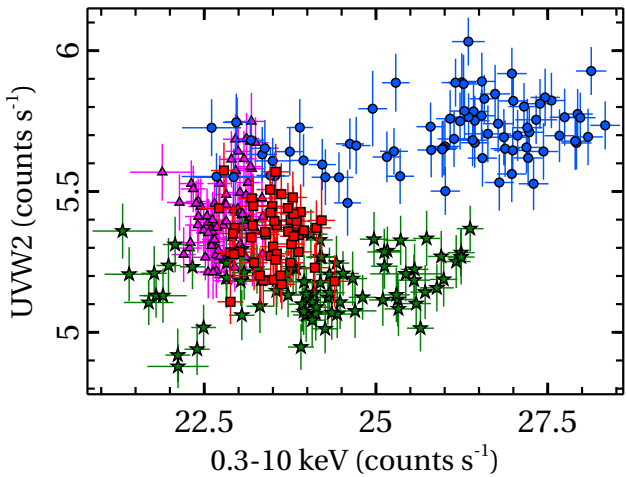}
\caption{The UVW2 count rate is plotted as a function of the X-ray ($0.3-10$\keV{}) count rate, implying a lack of correlation between the UV and X-ray bands at zero time-lag. The blue circle, red square, green star and magenta triangle represent data for the 2014a, 2014b, 2014c and 2014d observations, respectively. The time bin size used is 1\ks{}.}
\label{xray_w2}
\end{figure}

\begin{figure*}
\includegraphics[width=2\columnwidth]{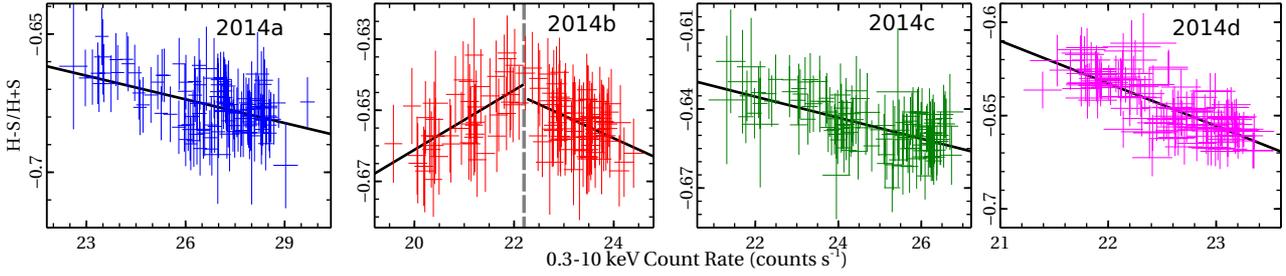}
\caption{The X-ray hardness ratio, H$-$S/H+S is plotted as a function of the X-ray (0.3$-$10\keV{}) count rate showing a softer-when-brighter trend except for 2014b where both softer (above $\sim22.1$~cts~s$^{-1}$) and harder (below $\sim22.1$~cts~s$^{-1}$) when brighter trends are observable. The solid black line shows the best-fitting linear model.}
\label{hr_flux}
\end{figure*}

\begin{table*}
\centering
\caption{The best-fitting parameters for the linear model fitting to the HID. ${\rm CR(min, max)}$ represents the 0.3$-$10\keV{} minimum and maximum background subtracted count rates of individual data used for fitting.}
\begin{center}
\scalebox{0.95}{%
\begin{tabular}{ccccccccc}                        
\hline       
Obs. Id  & ${\rm CR(min, max)}$ & Gradient  & Intercept & $\chi^2$/d.o.f & Behaviour when brighter \\ [0.2cm]
\hline 
 2014a & (22.6, 29.7) & $-0.00285\pm0.0005$ & $-0.6\pm0.01$  & 0.55  & Softer   \\ [0.2cm]
 2014b & (19.6, 22.1 & $0.0084\pm0.0017$  & $-0.83\pm0.03$  & 0.49 & Harder  \\ [0.2cm]
 2014b & (22.1, 24.4) & $-0.0064\pm0.0014$ & $-0.50\pm0.03$  & 0.35 & Softer  \\ [0.2cm]
 2014c & (21.3, 26.7) & $-0.004\pm0.0004$ & $-0.55\pm0.01$   & 0.49 & Softer   \\ [0.2cm]
 2014d &(21.4, 23.35) & $-0.023\pm0.002$   & $-0.13\pm0.04$  & 1.0 &  Softer  \\ [0.2cm]
\hline 
\end{tabular}}
\end{center}
\label{table5}           
\end{table*}

\begin{figure}
\includegraphics[width=\columnwidth]{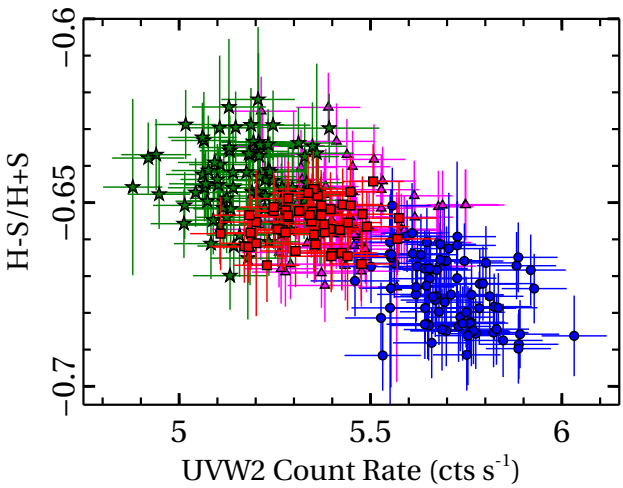}
\caption{The X-ray hardness ratio, H$-$S/H+S versus the UVW2 count rate showing a softer-when-brighter trend. The blue circle, red square, green star and magenta triangle represent data for 2014a, 2014b, 2014c and 2014d, respectively.}
\label{hr_w2}
\end{figure}

\begin{figure}
\includegraphics[width=\columnwidth]{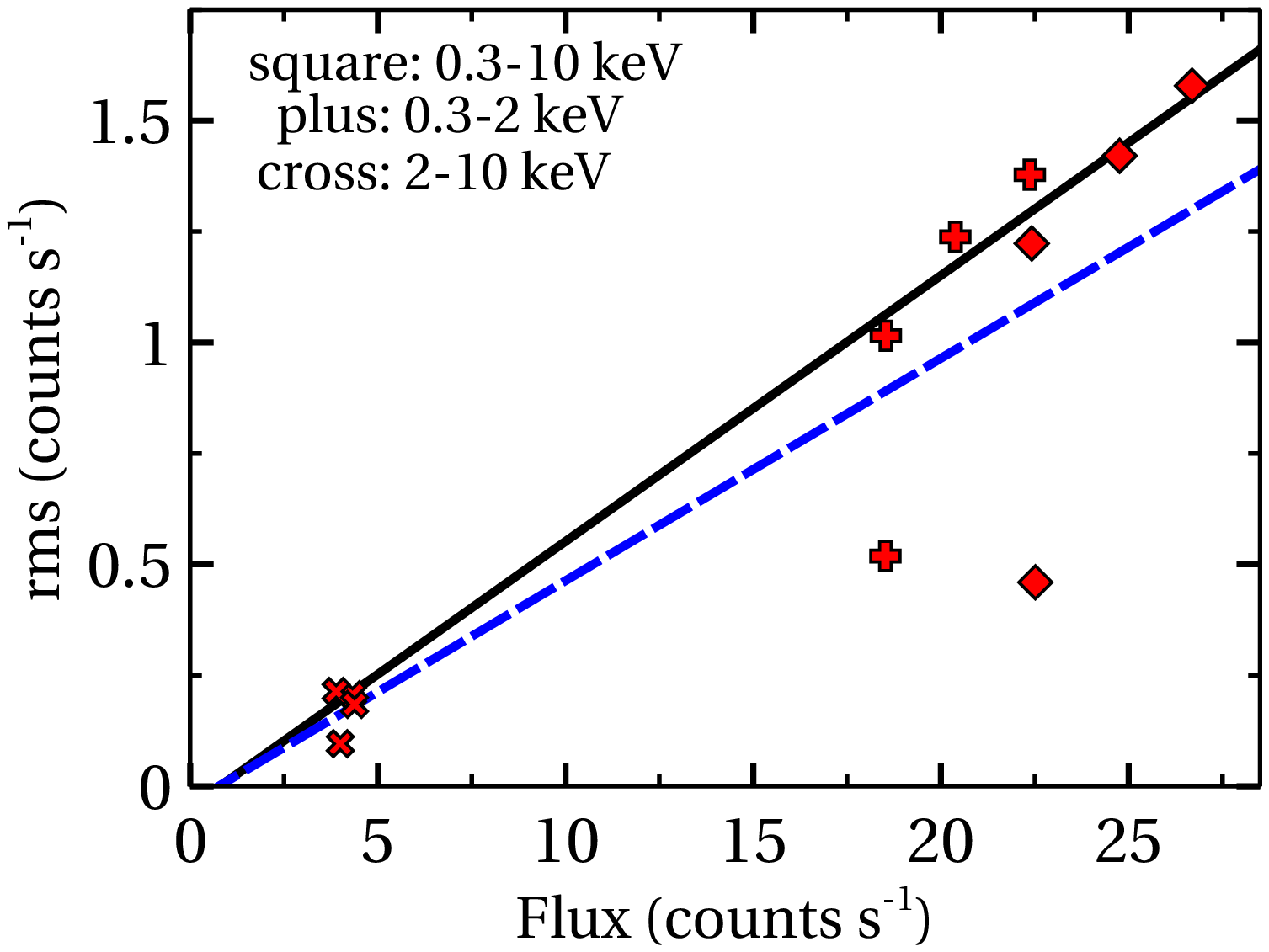}
\caption{The rms-flux relation for Ark~120 derived from the background subtracted EPIC-pn light curves in three energy bands: 0.3$-$10\keV{} (square), 0.3$-$2\keV{} (plus) and 2$-$10\keV{} (cross). The black solid line is the best-fitting linear model to the data excluding 2014d and the blue dashed line represents the deviation from the linear relation after including 2014d. The $1\sigma$ error bar is smaller than the marker size.}
\label{rms_flux}
\end{figure}

\section{Timing Analysis}

\subsection{Light Curve and Hardness Ratio}
Ark~120 showed moderate short-term variability during the 2014 \xmm{} observations. The top panels in Figure~\ref{lc} show the  background subtracted, 0.3$-$10\keV{} EPIC-pn light curves with the time bin size of 1\ks{}. A visual inspection of the light curve indicates that Ark~120 underwent a gradual flux variability by a factor of $\sim1.3$, $\sim1.25$, $\sim1.25$ and $\sim1.1$ within the 2014a, 2014b, 2014c and 2014d observations, respectively. The amplitude of the observed variations are of $\sim26.5$~per~cent, $\sim21.5$~per~cent, $\sim21.9$~per~cent and $\sim8.7$~per~cent of the mean count rate on $\sim120$\ks{} timescales derived from the 2014a, 2014b, 2014c and 2014d EPIC-pn light curves, respectively. The fractional root-mean-squared (rms) variability amplitudes in the full band (0.3$-$10\keV{}) are $F_{\rm var, X}=(5.92\pm0.08$)~per~cent, ($5.46\pm0.08$)~per~cent, ($5.74\pm0.08$)~per~cent and ($2.04\pm0.08$)~per~cent for 2014a, 2014b, 2014c and 2014d, respectively. The $1\sigma$ error on $F_{\rm var}$ was computed in conformity with \citet{va03}. In the bottom panel of Fig.~\ref{lc}, we have shown the simultaneous UVW2 light curves of Ark~120 with a time resolution of 1\ks{} extracted from the OM/fast mode observation. The fractional rms variability amplitudes in the UVW2 band are $F_{\rm var, UV}=(0.93\pm0.3$)~per~cent, ($0.97\pm0.4$)~per~cent, ($1.17\pm0.29$)~per~cent and ($1.67\pm0.24$)~per~cent for 2014a, 2014b, 2014c and 2014d, respectively.

To study the energy dependence of X-ray variability, we extracted the time series in two energy bands: soft (S=0.3$-$2\keV{}) and hard (H=2$-$10\keV{}). Figure~\ref{lc_hr} shows the background subtracted soft (top panel) and hard (middle panel) X-ray light curves of Ark~120 with the time binning of 1\ks{}. The pattern of variability between the soft and hard X-ray bands is similar except for 2014d where the trend is acutely different. The percentage of fractional variability in the soft band exceeds the hard band variability except for 2014b where the soft and hard bands have comparable variability amplitude. The mean count rate, the absolute rms amplitude and the fractional rms variability amplitude in the full, soft and hard bands are listed in Table~\ref{table3}. 

In the bottom panel of Fig.~\ref{lc_hr}, we have shown the time series of the hardness ratio which is defined by H$-$S/H+S. To investigate the time variability of the hardness ratio, we fitted a \textsc{constant} model to all four 2014 data sets which provided a statistically unacceptable fit with $\chi^{2}$/d.o.f = 1249/485, indicating the presence of a significant spectral variability over the period of $\sim7.5$~days. Thus the X-ray time series and hardness ratio analysis imply that the source is variable in flux as well as in spectral shape.

\subsection{Flux$-$Flux Analysis}
To investigate the connection between the soft and hard X-ray bands, we derived the H (=2$-$10\keV{}) vs S (=0.3$-$2\keV{}) flux$-$flux plots with the time bin size of 1\ks{}, which are shown in Figure~\ref{flux_flux}. The flux$-$flux analysis is a model-independent approach to probe the spectral variability and has been successfully applied in a number of Galactic sources and Seyfert galaxies (e.g. Cyg~X-1: \citealt{ch01}, MCG-6-30-15: \citealt{ta03}). First, we fitted a linear model of the form, $\rm H=m\times\rm S+c$ to the data which resulted in a good fit to all four 2014 observations except for 2014b. In order to test whether the pivoting of the primary emission is responsible for variability, we fitted a power-law without constant component to the data which provided an equally good fit except for 2014b where we found that a quadratic function of the form, $\rm H=a\times\rm S^{2}+b\times\rm S+c$ described the data better with $\chi^{2}$/d.o.f = 150.6/119. In the case of 2014d, the soft and hard band count rates are anti-correlated with the Spearman rank correlation coefficient of $\sim-0.5$ and null hypothesis probability of $\sim5.9\times10^{-9}$. The 2014 Flux$-$Flux plots of Ark~120 can be reasonably fitted by both the linear and power-law models except for 2014b where more complex modelling is required by the data. The results of the Flux$-$Flux modelling are listed in Table~\ref{table4}. We also investigated the variability relation between the UVW2 and X-ray bands. Figure~\ref{xray_w2} shows the variation of the UVW2 count rate as a function of the full band ($0.3-10$\keV{}) X-ray count rate. To test for any UV/X-ray correlation, we calculated the Spearman rank correlation coefficient between the UVW2 and X-ray count rates for all four 2014 observations. The estimated correlation coefficient is $\rho\simeq0.3$ which indicates that the X-ray and UV emission are not significantly correlated at zero time-lag. The lack of a formal correlation could be due to a sampling problem which arises when one section of the data is much more heavily sampled than another. There is a lot of scatter in the data and most of the points are concentrated in the bottom left region of Fig.~\ref{xray_w2}, which dominate the correlation.

\subsection{Hardness$-$Intensity Diagram}
The study of the hardness ratio (HR) versus total X-ray flux is an important model-independent approach which is used to determine the spectral evolution of the source and its association with the observed flux variability. Figure~\ref{hr_flux} shows the hardness$-$intensity diagram (HID) for all four EPIC-pn observations of Ark~120. The source showed a decrease in spectral hardness with increasing flux within the 2014a, 2014c and 2014d observations, although the HID of the 2014b observation is found to be an inverted-crescent shaped with two opposite trends below and above the mean count rate of $\sim22.1$~cts~s$^{-1}$. In order to numerically quantify whether Ark~120 is harder- or softer-when-brighter, we fitted the HID of the source with a linear model of the form ${\rm HR}=a \times{\rm CR}_{0.3-10}+b$. The best-fitting model parameters are listed in Table~\ref{table5}; the best-fitting model is shown as the solid black line in Fig.~\ref{hr_flux}. The modelling of the HID reveals the softer-when-brighter behaviour of Ark~120 during the 2014a, 2014c and 2014d observations. The only exception is the 2014b data where Ark~120 had the lowest flux during the first $\sim46$\ks{} of the observation and then the source switched from the harder-when-brighter to softer-when-brighter behaviour at a count rate of $\sim22.1$~cts~s$^{-1}$. In Figure.~\ref{hr_w2}, we have also shown the X-ray hardness ratio versus UVW2 count rate. The estimated Spearman rank correlation coefficient between the X-ray hardness and UVW2 count rate is $\sim-0.7$ with the null hypothesis probability of $p\sim2\times10^{-40}$. Thus we conclude that the source is brighter in the UVW2 band when it is softer in X-rays.

\subsection{The rms$-$flux relation}
We also derived the rms$-$flux relation which represents the behaviour of the absolute rms variability amplitude as a function of the X-ray flux and is commonly observed to be linear in X-ray binaries and AGN with a large range of black hole masses \citep{ut01,ga04,ut05}. The absolute rms variability amplitude is defined by the sample variance minus the uncertainties due to Poisson noise \citep{va03}:

\begin{equation}
\sigma_{\rm rms}^2=S^2-\overline{\sigma_{\rm err}^2}
\label{eq1}
\end{equation}
where $S^2$ is the total variance of the time series of length $N$ and $\overline{\sigma_{\rm err}^2}$ is the mean squared error which is defined by

\begin{equation}
\overline{\sigma_{\rm err}^{2}}=\frac{1}{N}\sum\limits_{i=1}^{N}\sigma_{\rm err,i}^{2}.
\end{equation}
We calculated the absolute rms in the background subtracted EPIC-pn light curves of time bins 1\ks{} using equation~(\ref {eq1}) in three different energy bands: full (0.3$-$10\keV{}), soft (0.3$-$2\keV{}) and hard (2$-$10\keV{}). The rms$-$flux relation for Ark~120 is shown in Figure~\ref{rms_flux}. To test the linearity of the relationship, we fitted a linear model of the form $y=mx+c$, which described the data well excluding 2014d. The best-fitting values of the slope and intercept are $m=0.06\pm0.002$ and $c=-0.047\pm0.021$, respectively. The black solid line is the best-fitting linear model to the data excluding 2014d. The inclusion of the 2014d data causes a slight deviation from the best-fitting linear relation due to its much less variability and is shown as the blue dashed line Fig.~\ref{rms_flux}.

\subsection{Fractional rms Spectral Modelling}
\begin{figure*}
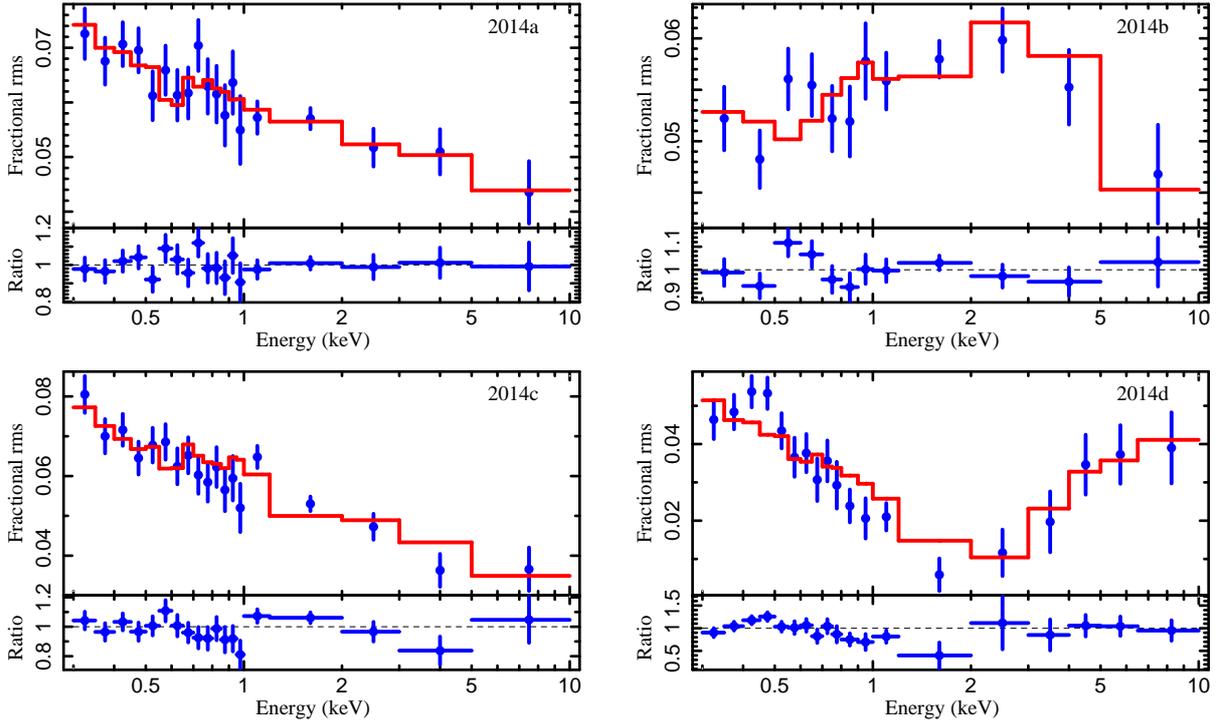

\includegraphics[scale=0.4,angle=-90]{fig10a.ps}
\includegraphics[scale=0.4,angle=-90]{fig10b.ps} 
\includegraphics[scale=0.4,angle=-90]{fig10c.ps}
\includegraphics[scale=0.4,angle=-90]{fig10d.ps}
\caption{The 2014 \xmm{}/EPIC-pn frequency averaged ($\sim 8-500\times10^{-6}$\hz{}) rms spectra, the best-fitting rms spectral model and the data-to-model ratio. The best-fitting model consists of two constant reflection components and two variable components: a soft excess with variable luminosity and a primary emission with variable normalization and spectral index.}
\label{rms}
\end{figure*}

To quantify the variability of different spectral parameters and the relationship between them, we derived and modelled the fractional rms spectrum of Ark~120. First, we generated background subtracted EPIC-pn light curves in a number of energy bands with a time resolution of $\Delta t=1$\ks{}. Then we calculated the fractional rms variability amplitude $F_{\rm var}$ and its $1\sigma$ error in each light curve using the formula given in \citet{va03}. The frequency averaged ($\sim 8-500\times10^{-6}$\hz{}) fractional rms spectra of Ark~120 are shown in Figure~\ref{rms}. The source showed a decrease in fractional variability with energy for 2014a and 2014c. However, we found noticeably different rms spectra in the case of 2014b and 2014d. The 2014b variability spectrum showed an increase in fractional rms with energy up to $\sim2$\keV{} and then it started decreasing. For 2014d, the fractional variability decreases with energy until $\sim2$\keV{} and then it begins to rise with energy.

The fractional rms variability spectrum can distinguish the variable spectral components present in the mean spectrum and therefore can be considered as a useful tool to probe the spectral variability of AGN (\citealt{mi07,fa12,ma16,md17}). In order to identify the spectral components responsible for the observed X-ray variability and to compute the fractional variability of variable spectral parameters, we modelled the fractional rms variability spectra of Ark~120. First we generalized the equation~(4) of \citet{ma16} for $n$ numbers of variable spectral parameters ($p_{1},p_{2},...,p_{n}$) of the mean spectrum $f(E,p_{1},...,p_{n})$ consisting of $m$ numbers of spectral components ($f_{1},f_{2},..f_{m}$). Therefore the expression for the fractional rms variability spectrum $F_{\rm var}(E)$ can be written as
\begin{equation}
F_{\rm var}(E)=\frac{\sqrt{<(\Delta f(E,p_{1},...,p_{n}))^{2}>}}{f(E,p_{1},...,p_{n})}
\end{equation}

where \begin{equation}
f(E,p_{1},...,p_{n})=f_{1}(E,p_{1},...,p_{n})+...+f_{m}(E,p_{1},...,p_{n})
\label{eq2}
\end{equation}

and \begin{equation}
\Delta f(E,p_{1},...,p_{n})=f(E,p_{1}+\Delta p_{1},...,p_{n}+\Delta p_{n})-f(E,p_{1},...,p_{n})
\label{eq3}
\end{equation} 
The equation~(\ref{eq2}) represents the best-fitting mean spectral model which consists of a thermally Comptonized primary continuum ($f_{\rm nth}$), a blurred reflection ($f_{\rm blur}$), a distant reflection ($f_{\rm distant}$) and an intrinsic disc Comptonized component ($f_{\rm optxagn}$). Mathematically, the expression for the best-fitting mean spectral model excluding the Galactic absorption can be written as
\begin{equation}
f(E)=f_{\rm nth}(E)+f_{\rm blur}(E)+f_{\rm distant}(E)+f_{\rm optxagn}(E)
\end{equation}
The equation~(\ref{eq3}) represents variations in the mean spectrum arising due to variable spectral components. If the observed X-ray variability is due to variations in the normalization ($p_{1}\equiv N_{\rm nth}$) and slope ($p_{2}\equiv\Gamma$) of the primary emission (\textsc{nthcomp}) and the luminosity ($p_{3}\equiv \log L_{\rm s}$) of the soft excess emission (\textsc{optxagnf}) then the equation~(\ref{eq3}) can be written as
\begin{small}
\begin{equation}
\Delta f(E,N_{\rm nth},\Gamma,L_{\rm s})=\Delta f_{\rm nth}(E,N_{\rm nth},\Gamma)+\Delta f_{\rm optxagn}(E,\log L_{\rm s})
\end{equation}
\end{small}
where
\begin{small}
\begin{equation}
\Delta f_{\rm nth}(E,N_{\rm nth},\Gamma)=f_{\rm nth}(N_{\rm nth}+\Delta N_{\rm nth},\Gamma+\Delta \Gamma)-f_{\rm nth}(N_{\rm nth},\Gamma)
\label{eq4}
\end{equation}
\end{small}
and
\begin{small}
\begin{equation}
\Delta f_{\rm optxagn}(E,\log L_{\rm s})=f_{\rm optxagn}(\log L_{\rm s}+\Delta \log L_{\rm s})-f_{\rm optxagn}(\log L_{\rm s})
\label{eq5}
\end{equation}
\end{small}
Therefore the expression for the fractional rms can be written as
\begin{small}
\begin{equation}
 F_{\rm var}=\frac{\sqrt{<(\Delta f_{\rm nth}(E,N_{\rm nth},\Gamma)+\Delta f_{\rm optxagn}(E,\log L_{\rm s}))^2>}}{f_{\rm nth}(E)+f_{\rm blur}(E)+f_{\rm distant}(E)+f_{\rm optxagn}(E)}
\label{eq6}
\end{equation}
\end{small}
We then expanded the first term on the right-hand side (R.H.S) of equation~(\ref{eq4}) and (\ref{eq5}) in a Taylor series around the variable parameters ($N_{\rm nth}$, $\Gamma$) and $\log(L_{\rm s}/L_{\rm E})$, respectively and then ignored higher order (from the second order derivatives onward) terms. The Taylor series expansion also provides the correlation coefficient, $\alpha$ between these two parameters, $\Delta N_{\rm nth}$ and $\Delta \Gamma$ of the primary spectral component (\textsc{nthcomp}). We then constructed the fractional rms spectral model (equation \ref{eq6}) in \textsc{S-Lang}~V2.3.0 and implemented in \textsc{ISIS}~V.1.6.2-32 \citep{ho00} as a local model.

\begin{table*}
 \centering
 \caption{The best-fitting fractional rms spectral (0.3$-$10\keV{}) model parameters of Ark~120. Here $\Delta N_{\rm nth}$, $\Delta \Gamma$ represent the variations in the normalization and spectral index of the thermally Comptonized primary continuum, $\alpha$ is the correlation between $\Delta N_{\rm nth}$ and $\Delta \Gamma$. $\Delta L_{\rm s}$ represents the variation in the luminosity of the disc Comptonized soft excess continuum.}
\begin{center}
\scalebox{1.0}{%
\begin{tabular}{ccccccc}
\hline 
Variable & Model & Parameter  & 2014a & 2014b & 2014c & 2014d  \\
Component &     & & & & & \\
\hline 
Primary Continuum  & \textsc{nthcomp} & $\frac{\Delta N_{\rm nth}}{N_{\rm nth}}$~(per~cent) &$7.8^{+0.7}_{-0.7}$ & $8.4^{+0.6}_{-0.6}$ & $8.1^{+0.8}_{-0.9}$ & $1.5^{+1.1}_{-1.1}$  \\ [0.2cm]             
 & & $\frac{\Delta \Gamma}{\Gamma}$~(per~cent) & $1.4^{+0.6}_{-0.6}$ & $0.6^{+0.4}_{-0.3}$ & $1.7^{+0.4}_{-0.5}$ & 0.8$^{+0.2}_{-0.3}$ \\ [0.2cm]    
  & & $\alpha$ & 0.8$^{+0.2}_{-0.1}$ & 0.88$^{+0.12}_{-0.18}$ & 0.9$^{+0.05}_{-0.07}$ & 1$^{+0}_{-0.24}$ \\ [0.2 cm]      
Soft Excess &\textsc{optxagnf} & $\frac{\Delta L_{\rm s}}{L_{\rm s}}$~(per~cent)  & $3.5^{+0.2}_{-0.2}$ & $2.4^{+0.4}_{-0.4}$ & $4.1^{+0.2}_{-0.2}$ & $2.9^{+0.2}_{-0.2}$ \\ [0.2cm]
 & & C-stat/d.o.f & 9.2/15 & 11.1/8 & 20.6/15 & 27.2/15 \\ [0.2cm]
\hline
\end{tabular}}
\end{center}
\label{table6}
\end{table*}

Initially, we fitted the $0.3-10$\keV{} fractional rms spectra with the primary continuum (\textsc{nthcomp}) model having only variable normalization $\Delta N_{\rm nth}$. This resulted in a statistically unacceptable fit with C-stat/d.o.f = 725.6/18, 280/11, 839.3/18 and 368.4/18 for 2014a, 2014b, 2014c and 2014d respectively. Then we introduced the slope variation ($\Delta \Gamma$) of the primary continuum and also the correlation coefficient ($\alpha$) between $\Delta N_{\rm nth}$ and $\Delta \Gamma$, which improved the fit to C-stat/d.o.f = 268.3/16 ($\Delta C$=$-$457 for 2 d.o.f), 41.3/9 ($\Delta C$=$-$239 for 2 d.o.f), 639.3/16 ($\Delta C$=$-$200 for 2 d.o.f) and 146.2/16 ($\Delta C$=$-$222 for 2 d.o.f) for 2014a, 2014b, 2014c and 2014d, respectively. However, the variable primary emission (\textsc{nthcomp}) model underestimated the low energy ($0.3-1$\keV{}) variability which is basically driven by the soft excess. Therefore we introduced variability in the soft excess luminosity, $\Delta \log L_{\rm s}$ of the \textsc{optxagnf} model which provided a significant improvement in the fitting with C-stat/d.o.f = 9.2/15 ($\Delta C$=$-$259), 11.1/8 ($\Delta C$=$-$30.2), 20.6/15 ($\Delta C$=$-$618.7) and 27.2/15 ($\Delta C$=$-$119) for 2014a, 2014b, 2014c and 2014d, respectively. The best-fitting model parameters are listed in Table~\ref{table6}. The best-fitting rms model for all four observations is shown as solid red lines in Fig.~\ref{rms}. Thus the modelling of the X-ray (0.3$-$10\keV{}) variability spectra suggests the presence of two constant reflection components, a variable soft excess and a variable primary continuum in Ark~120.

\begin{figure}
\includegraphics[width=\columnwidth]{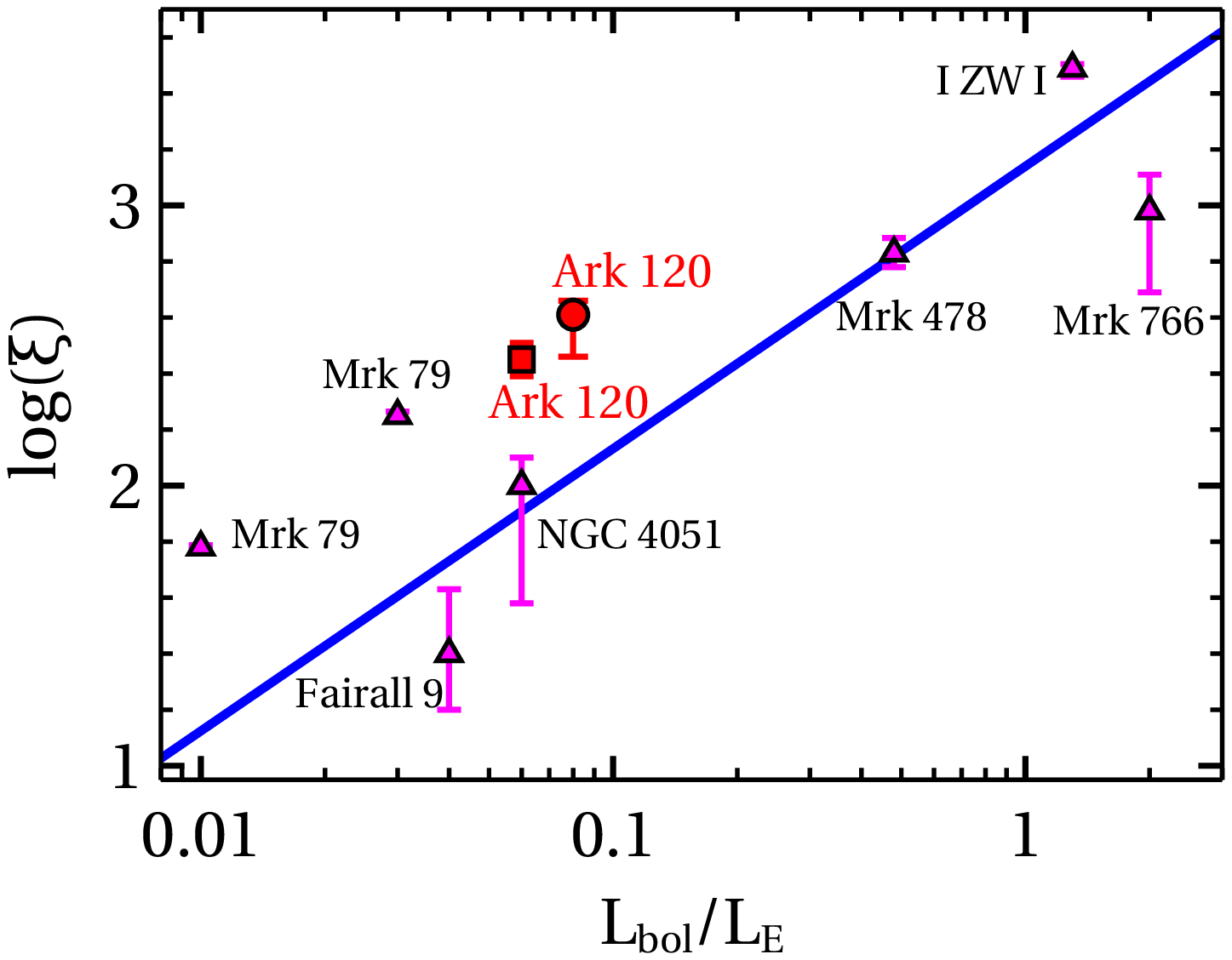}
\caption{Dependence of disc ionization parameter ($\log\xi$) on the Eddington ratio ($L_{\rm bol}/L_{\rm E}$), for a number of Seyfert galaxies. The solid line represents the linear relation [$\log\xi=1.008\log(L_{\rm bol}/L_{\rm E})+3.14$] from \citet{is90}. The square and circle represent Ark~120 from \citet{na11} and this work, respectively.}
\label{logxi}
\end{figure}

\begin{figure}
\includegraphics[width=\columnwidth]{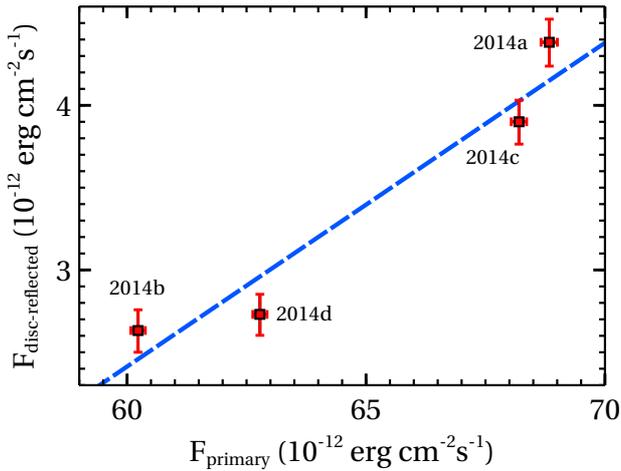}
\caption{The illuminating continuum flux vs. disc reflected flux in the full ($0.3-10$\keV{}) band, obtained from four observations in 2014.}
\label{pl_xil}
\end{figure}

\begin{figure}
\includegraphics[width=\columnwidth,angle=-0]{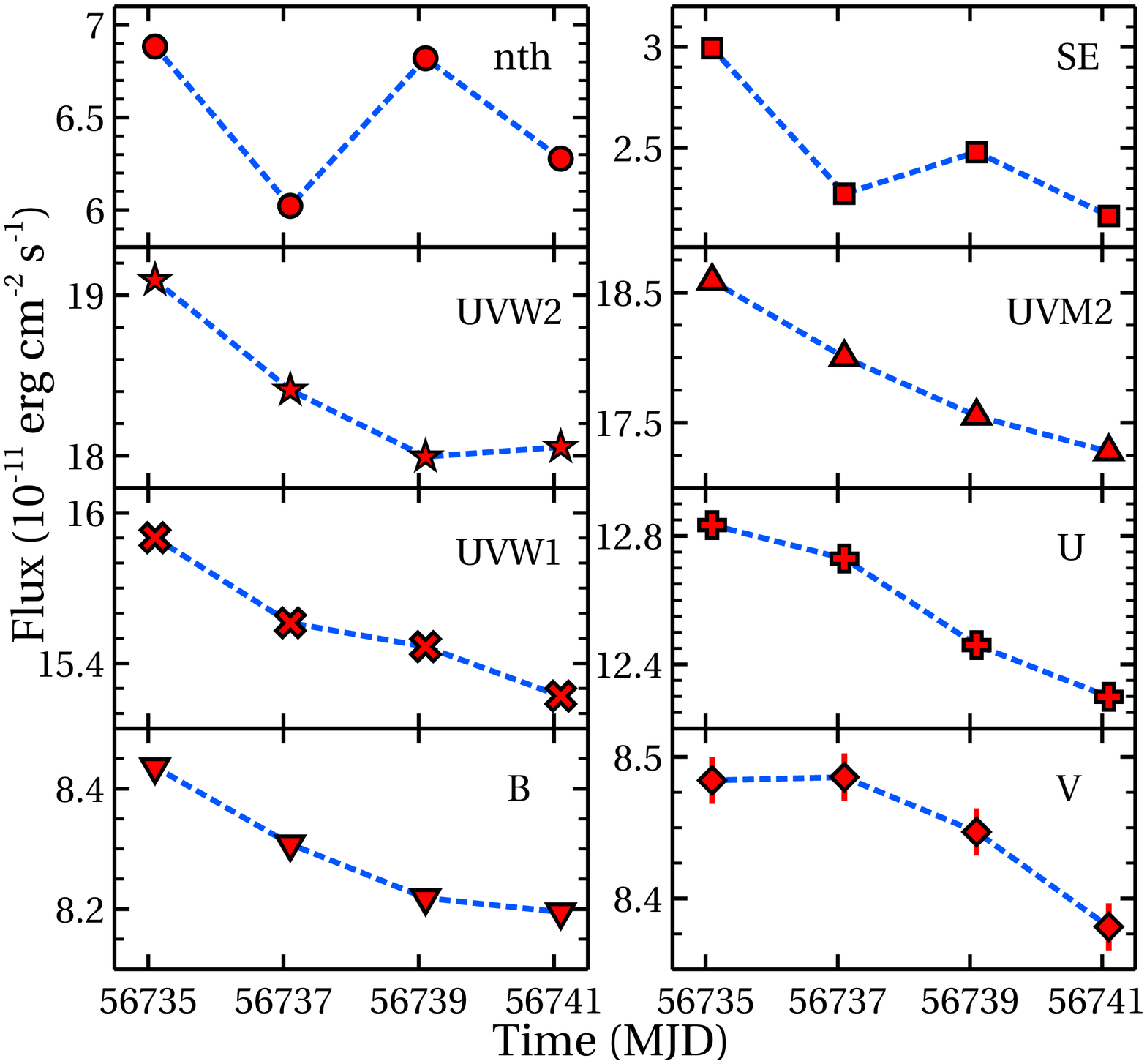}
\caption{Time evolution of the primary continuum (nth) flux, the soft excess (SE) flux and the de-reddened UV (UVW2, UVM2, UVW1), optical (U, B, V) flux during the course of the 2014 observations. The $1\sigma$ error bar is smaller than the marker size.}
\label{flux_obs}
\end{figure}

\section{Discussion and Conclusions}

\subsection{Summary}
We present results from a long ($\sim486$\ks{}) \xmm{}/EPIC-pn and OM analysis of the bare Seyfert~1 galaxy Ark~120 observed during 2014 March. We investigated the small-amplitude, short-term variability in the source through a detailed analysis of the X-ray spectral-timing properties and UV/X-ray connection over the observed $\sim1$-week period. We also distinguished between the variable components (primary continuum and soft excess) using the fractional rms spectral modelling. We summarize the main results of our work below.
\begin{enumerate}
\item The X-ray (0.3$-$10\keV{}) spectrum of Ark~120 is well described by a thermally Comptonized primary continuum with two (one blurred and one distant) reflection components and a warm Comptonization component for the soft X-ray continuum. The lower limit on the inner disc radius estimated from the blurred reflection model is $\sim38.8 r_{\rm g}$, $\sim25.1 r_{\rm g}$, $\sim30.1 r_{\rm g}$ and $\sim12.7 r_{\rm g}$ for 2014a, 2014b, 2014c and 2014d, respectively, which are in close agreement with the results obtained by \citet{na16}. The coronal radius determined from the joint fitting of four X-ray spectra is $r_{\rm corona}=7.6^{+0.1}_{-0.1}r_{\rm g}$.

\item Ark~120 was observed in a moderately variable state with the fractional X-ray (0.3$-$10\keV{}) and UVW2 variability amplitudes to lie in the range of $\sim2-6$~per~cent and $\sim1-2$~per~cent, respectively on timescales of $\sim7.5$~days. 

\item The joint spectral fitting of all four observations (2014a, 2014b, 2014c and 2014d) revealed a significant variability in the soft excess luminosity, primary power-law normalization and photon index, although there was no noticeable variation in the electron temperature ($kT_{\rm {SE}}=0.37\pm0.01$\keV{}) of the optically thick, warm corona.

\item The fractional variability amplitude in the soft band (0.3$-$2\keV{}) is higher than the hard band (2$-$10\keV{}) which is indicative of the presence of multiple spectral components varying differently, while for the second observation (2014b), the fractional rms amplitude variation in these two bands is found to be similar.

\item Ark~120 exhibits more variability when brighter which is consistent with the linear rms-flux relation observed in accreting objects over a wide range in black hole mass. However, the variability in 2014d was much less than the other three observations and hence causes a slight deviation from the linear relation (see Fig.~\ref{rms_flux}). 

\item The soft (0.3$-$2\keV{}) and hard (2$-$10\keV{}) band count rates are found to be correlated with each other for the 2014a, 2014b and 2014c observations. However, for the 2014d observation, we found a moderate anti-correlation between these two bands (see Fig.~\ref{flux_flux}). The hard band positive offset observed in all four flux--flux plots can be interpreted as a corroboration for the presence of a constant component at higher energies, which is in agreement with the non-variability of the reflection components (iron emission complex) as inferred from the X-ray variability spectral modelling. A similar behaviour was also observed in another Seyfert~1 galaxy MCG-6-30-15 \citep{mi07}.

\item The fractional rms variability decreases with energy for 2014a and 2014c. However, the X-ray rms variability spectra for 2014b and 2014d are found to be inverted-crescent and crescent shaped, respectively (see Fig.~\ref{rms}). The modelling of the X-ray (0.3$-$10\keV{}) variability spectra confirmed the presence of constant reflection components, a variable disc Comptonized soft excess emission and a variable illuminating continuum (see Table~\ref{table6}).

\item The hardness ratio analysis shows a softer-when-brighter behaviour for Ark~120 during the 2014 observations. However, we found a dual nature of the hardness$-$intensity diagram (HID) during the 2014b observation where the source evolves from harder-when-brighter to softer-when-brighter behaviour at the mean count rate of $\sim22.1$~cts~s$^{-1}$ (see Fig.~\ref{hr_flux}). A detailed discussion of the peculiar HID including flux resolved rms spectrum for the second observation is deferred to a future paper. 

\end{enumerate}

\subsection{Primary Continuum Variability}
The energy-dependent X-ray variability of Ark~120 showed strong modulation during 2014 March as observed in Fig.~\ref{rms}. Our spectral and timing analyses suggest that the observed moderate variability is predominantly due to the primary continuum which is variable both in flux and spectral shape. The variations in the normalization and spectral index of the hot coronal emission are correlated with each other by $\sim70-100$~per~cent. The softening of the source with increasing $0.3-10$\keV{} flux is consistent with that observed from the radio-quiet AGN (e.g. \citealt{vau01,pa07,em11}). This type of variability is generally attributed to the thermal Comptonization where an increase in the seed photon flux cools down the hot corona, resulting in a steeper primary emission. We also found a harder-when-brighter behaviour of Ark~120 in the beginning (first 46\ks{}) of the 2014b observation during which the source was in the lowest flux state among all four observations. Such a behaviour is generally observed in blazars (e.g. \citealt{kr04,gl06,ma16}) or in low luminosity AGN (e.g. \citealt{em12,co16}). The intrinsic X-ray (0.3$-$10\keV{}) luminosities of Ark~120 are $2.42\times10^{44}$~erg~s$^{-1}$, $2.0\times10^{44}$~erg~s$^{-1}$, $2.26\times10^{44}$~erg~s$^{-1}$ and $2.1\times10^{44}$~erg~s$^{-1}$ for 2014a, 2014b, 2014c and 2014d respectively, while for the first 46\ks{} of the 2014b observation, the source luminosity is estimated to be $1.9\times10^{44}$~erg~s$^{-1}$ which is about 1.3 times less than the 2014a observation. The dominant variable component appears to be the primary continuum (see Table~\ref{table6}) during the 2014b observation, thus the dual nature of the HID is likely to be caused by intrinsic variations of the hot corona.

To investigate the disc/corona interaction in the inner regions of the accretion disc, we studied the dependence of disc ionization state ($\log\xi$) on the Eddington ratio ($L_{\rm bol}/L_{\rm E}$) which is considered as a probe to test the $\alpha$-disc accretion theory \citep{sh73}. According to the standard $\alpha$-disc model, the disc ionization parameter is expected to be correlated with the bolometric luminosity of AGN \citep{is90,ba11}. We estimated the Eddington ratio of Ark~120 by applying the 2$-$10\keV{} bolometric correction ($\kappa=17.8$, \citealt{vasetal09}) to the X-ray (2$-$10\keV{}) luminosity. By definition, $\kappa=L_{\rm bol}/L_{\rm (2-10\keV{})}$. We calculated the unabsorbed 2$-$10\keV{} flux from the combined time-averaged spectrum using the convolution model \textsc{cflux} in \textsc{xspec}. Thus the bolometric luminosity is estimated to be $L_{\rm bol}\approx 1.6\times10^{45}\rm~erg~s^{-1}$. For $M_{\rm BH}\sim 1.5\times10^{8} M_{\odot}$ \citep{pe04}, the Eddington luminosity is $L_{\rm Edd}=20.7\times10^{45}\rm~erg~s^{-1}$ and hence the Eddington ratio for Ark~120 is $\dot{m}_{\rm Edd}=L_{\rm bol}/L_{\rm E}\approx0.08$ which is around 1.6 times higher than the Eddington rate ($\dot{m}_{\rm Edd}\sim0.05$) estimated by \citet{vas07} using the \asca{} X-ray data. The discrepancy in the Eddington ratio is due to variations in the intrinsic 2$-$10\keV{} source luminosity between 1994 \asca{} and 2014 \xmm{} observations. Figure~\ref{logxi} shows the dependence of disc ionization parameter ($\xi$) on the Eddington ratio for a number of Seyfert galaxies \citep{ba11} including Ark~120 (\citealt{na11} and this work). The solid line represents the linear relation [$\log\xi=1.008\log(L_{\rm bol}/L_{\rm E})+3.14$] from \citet{is90}. The departure of Ark~120 from the predicted linear $\log\xi-L_{\rm bol}/L_{\rm E}$ relation can be attributed to the changing coronal power assuming a constant radiative efficiency of the accretion disc \citep{ba11}, which is further supported by the presence of a variable coronal emission in the X-ray rms spectra of Ark~120.

The nature of the hard band variability in Ark~120 is complex and the fractional rms variability pattern is changing from observation-to-observation (see Fig.~\ref{rms}) which can be explained in terms of the interplay between the primary continuum (flux/spectral shape) and reflected emission from the accretion disc. In Figure~\ref{pl_xil}, we have shown the variation of the disc reflected flux as a function of the intrinsic continuum flux during the course of the 2014 observations. The approximately linear correlation between the primary continuum and disc reflected fluxes implies that the variation in the highly ionized disc reflected emission is driven by the changes in the illuminating flux and thus variability is intrinsic to the X-ray source. Therefore the complex hard X-ray variability of Ark~120 can be attributed to the X-ray emitting hot corona with a variable intrinsic luminosity. The fractional variations in the normalization of the incident continuum are $\sim7.8$~per~cent, $\sim8.4$~per~cent, $\sim8.1$~per~cent and $\sim1.5$~per~cent for 2014a, 2014b, 2014c and 2014d, respectively. Above $\sim2$\keV{}, the fractional rms decreases with energy for the first three observations. This behaviour is due to spectral steepening with flux possibly resulting from changes in the optical depth and/or electron temperature of the hot corona. A similar hard X-ray rms variability pattern has been observed in other Seyfert galaxies (e.g. IC~4329A: \citealt{br14} and NGC~4151: \citealt{kec15}). In the case of 2014d observation, the fractional rms variability above $\sim2$\keV{} increases with energy which is probably due to the hot coronal compactness variability with a minimal normalization variability of the hard Comptonized component
\citep{gi05}. From the X-ray variability spectral modelling, we found that the fractional variation in the normalization of the primary continuum for 2014d observation is only $\sim1.5$~percent which is much less (a factor of $\sim5$) compared to the other three observations (see Table~\ref{table6}).

\begin{figure}
\includegraphics[width=\columnwidth]{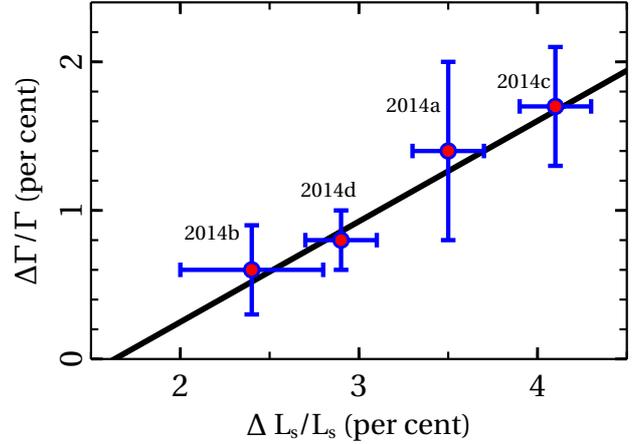}
\caption{The fractional variation in the spectral index, ($\frac{\Delta \Gamma}{\Gamma}$) of the primary continuum as a function of the luminosity variation ($\frac{\Delta L_{\rm s}}{L_{\rm s}}$) of the soft excess emission for all four observations.}
\label{gama_Ls}
\end{figure}

\subsection{Soft Excess Variability}
In all four observations, Ark~120 showed a strong soft X-ray excess emission which is described by the intrinsic disc Comptonization model, which requires a warm corona with the electron temperature of $T_{\rm SE}\sim0.36$\keV{} and the optical depth of $\tau\sim12$. The observed soft excess emission is variable within individual observations ($\sim1.4$~days) as well as between observations typically separated by $\sim1$~day. The X-ray rms spectral modelling implies that the observed soft excess variability in the $0.3-1$\keV{} band is caused by the luminosity variation of the warm corona. The fractional variations in the luminosity of the soft excess emitting warm corona are $\sim3.5$~per~cent, $\sim2.4$~per~cent, $\sim4.1$~per~cent and $\sim2.9$~per~cent for 2014a, 2014b, 2014c and 2014d, respectively. Although the soft excess flux is variable, we did not find any significant variations in the electron temperature and optical depth of the warm corona. The variability pattern of the soft excess flux appears to be correlated with both the intrinsic power-law flux and de-reddened UV (in particular UVW1) flux over the entire $\sim7.5$~days period (Figure~\ref{flux_obs}). Moreover, the variations in the soft excess luminosity and the primary continuum photon index are correlated with each other over the observed timescale (see Fig.~\ref{gama_Ls}), which further supports the softer-when-brighter behaviour of Ark~120. All these characteristics are in favour of Comptonization models where the soft excess and primary X-ray emission are produced through Compton up-scattering of the UV and UV/soft X-ray seed photons, respectively.

\subsection{UV/X-ray Connection}
The fractional UV variability amplitude in Ark~120 is lower (a factor of $\sim2-3$) compared to the X-ray variability which is expected since the intrinsic UV emission originates from the outer disc and/or broad-line region (BLR) \citep{re16}. For Ark~120, \citet{gl17} found that the X-rays led UV by $7.5\pm7$~days. The lack of a correlation between the UV and X-ray emission at zero lag may be due to a long reprocessing delay between these two bands or a sampling problem which is caused by the concentration of data points in a narrow range of the UVW2 vs. X-ray flux-flux plot (see Fig.~\ref{xray_w2}). We found that the UV emission from Ark~120 is anti-correlated with the X-ray hardness ratio (see Fig.~\ref{hr_w2}), which suggests that the UV seed photons from the accretion disc/BLR get Compton up-scattered in the corona and as the seed photon supply to the corona increases, the corona cools down and hence the observed X-ray spectrum becomes steeper. This behaviour further supports Comptonization scenario as a physical process responsible for the connection between the UV and X-ray emission from Ark~120 over the observed $\sim1$-week period. A similar UV flux/X-ray hardness ratio anti-correlation has been observed in a few other Seyfert galaxies (e.g. MCG-6-30-15: \citealt{are05} and NGC~5548: \citealt{mc14}).

The X-ray spectral softening of the source with the UV flux (Fig.~\ref{hr_w2}), as well as an identical variability trend between the UVW1/soft excess or UVW1/primary continuum flux over the observed $\sim7.5$~days timescales (Fig.~\ref{flux_obs}) indicate that both the soft excess emitting warm corona and primary X-ray emitting hot corona are powered by the Comptonization of the lower energy UV seed photons. The observed linear correlation between the fractional variations in the soft excess luminosity and spectral index of the primary continuum (Fig.~\ref{gama_Ls}) implies that the Comptonized soft X-ray photons from the warm corona can also act as seed photons for the production of the primary X-ray emission in the hot corona. However, distinguishing the spatial separation between the warm and hot coronae quantitatively is beyond the scope of this work. Our spectro-temporal analysis suggests that the observed energy-dependent variability of Ark~120 is a consequence of variations in the spectral shape and intrinsic luminosity of the primary X-ray emitting hot corona, as well as the luminosity of the soft excess emitting warm corona, both of which are driven by variations in the Comptonizing seed photon flux.

\section{Acknowledgements}
The authors thank the anonymous referee for valuable comments and suggestions which helped to improve the manuscript. LM is thankful to Poshak Gandhi, Ranjeev Misra and Julien Malzac for useful comments. IMcH thanks the Royal Society for support under a Royal Society Leverhulme Trust Senior Research Fellowship, and thanks STFC for support under grant ST/M001326/1. This research has made use of archival data of \xmm{} observatory through the High Energy Astrophysics Science Archive Research Center Online Service, provided by the NASA Goddard Space Flight Center. This research has made use of the NASA/IPAC Extragalactic Database (NED), which is operated by the Jet Propulsion Laboratory, California Institute of Technology, under contract with the NASA. Figures in this paper were made with the GUI scientific plotting package \textsc{veusz}.

\bsp	
\label{lastpage}
\end{document}